\documentclass[
11pt,reprint,
onecolumn,
amsmath,amssymb,
aps,
]{revtex4-2}

\usepackage{upgreek} 
\usepackage{physics} 
\usepackage{graphicx}
\usepackage{subfigure}

\usepackage[english]{babel}
\usepackage{graphicx}
\usepackage{dcolumn}
\usepackage{bm}
\usepackage{xcolor}
\usepackage{array}
\usepackage{tabularx}
\usepackage{ragged2e}  
\usepackage{multirow}
\usepackage[frak = euler]{mathalpha}
\usepackage[citecolor=red, colorlinks, linkcolor=blue, urlcolor=blue,]{hyperref}
\usepackage{tikz}
\usetikzlibrary{decorations.pathmorphing,patterns}
\usepackage[dvipsnames]{xcolor}
\usepackage{mathrsfs} 

\newcommand{\be}{\begin{equation}}
	\newcommand{\ee}{\end{equation}}
\newcommand{\bea}{\begin{eqnarray}}
	\newcommand{\eea}{\end{eqnarray}}

\begin{document}
	
	\title{Spreading and rupture dynamics of soluble surfactant-laden thin film flow down slippery incline in presence of external shear}
	
	\author{Dipankar Paul}
	\email{dipankarpaul3103@gmail.com}
	\affiliation{Department of Mathematics, SRM Institute of Science and Technology, Chennai, India}
	
	\author{Harekrushna Behera}
	\email{hkb.math@gmail.com}
	\affiliation{Center of Excellence for Ocean Engineering, National Taiwan Ocean University, Keelung 202301, Taiwan}

	\begin{abstract}
		The spreading and rupture of the local distribution of surfactant on a slippery inclined thin film flow in the presence of external shear is explored in this article. The surfactant can be adsorbed at the free surface or can be dissolved in the bulk. The surfactant concentrations are governed by advection and diffusion equations for the bulk as well as the interface.  Moreover, the adsorption-desorption rates of the bulk and interface surfactants are regulated by sorption kinetics rates. The van der Waals forces are considered for rupture dynamics. The lubrication approximation method is used to derive the evolution equations of film thickness and interface surfactant concentration. Also, the rapid vertical diffusion is considered for cross-sectional averaged concentration to derive the evolution equation of bulk surfactant concentration. Besides, van der Waals forces are taken into account to observe the rupture dynamics. Two different scenarios are considered for sorption kinetics rates: (i) rapid and (ii) slow in the case of the spreading phenomenon. Then, two cases are considered related to the distribution of the surfactant in case of rapid sorption kinetics. The slippery bottom helps more fluid to flow and the capillary ridge to gain height for rapid sorption kinetics, and the external shear force amplifies the thinning of the film. However, in the case of slow sorption kinetics, the adsorption-desorption takes place at a slow rate, and the transient period results in a reduced Marangoni gradient at the interface. This leads to a pulse-type character in the film thickness profile. The external shear force reduces the pulse height, whereas a slippery surface at the bottom increases the pulse height. On the other hand, van der Waals forces are considered to be the major factor behind the rupture mechanism. The linear stability analysis depicts that external shear force destabilizes the flow, but the slip parameter displays a dual effect based on the Bond number, capillary number, and Hamaker constant. 
	\end{abstract}
	
	\maketitle

	\section{Introduction} \label{sec:Introduction}
	
	If a certain amount of surfactant is deposited on a thin film flow, then surface tension gradients will create flow, which eventually spreads the surfactants throughout. This phenomenon is called Marangoni spreading. For a thin film fluid with insoluble surfactant-laden flow, Marangoni spreading can generate significant deformation of the fluid layer, and film height exhibits shock at the leading edge of the monolayer \cite{borgas1988monolayer, troian1990model, gaver1990dynamics}. On the other hand, if the surfactant is soluble, then film thickness demonstrates a pulse-like structure instead of a shock, with a steep slope at the upstream end\cite{jensen1993spreading}. This Marangoni spreading has a widespread application in the industrial sector, such as, ultra-clean technology \cite{matar2001models}, coating \cite{kim2016controlled, baumgartner2022marangoni}, oil spreading on the sea \cite{hoult1972oil}, encapsulation process \cite{koldeweij2019marangoni}, pattern deposition \cite{mouat2019liquid, kaplan2015dynamics} and nature, as surface replacement therapy \cite{grotberg2001respiratory}, drug delivery for airway infection \cite{stetten2018surfactant, ma2020fingering} etc. 
	
	The lubrication approximation method \cite{borgas1988monolayer, gaver1990dynamics} is one of the most widely used mechanisms to handle the spreading phenomena of thin film flow \cite{kovalchuk2016effect, warner2004fingering}, droplet \cite{ jalaal2021spreading, kubochkin2021surface, saiseau2022near} etc \cite{francca2024elasto}. This method allows studying wave propagation, flow of insoluble and soluble surfactant-laden flow, and flow over an incline, and has discovered some of the most interesting phenomena. Using this method, typically two nonlinear evolution equations are obtained involving film thickness $h(x,t)$ and surfactant concentration $\Gamma(x,t)$. These equations are then solved either numerically or analytically to comprehend the spreading dynamics. For insoluble surfactant-contaminated thin film flow, \citet{jensen1992insoluble} used lubrication theory and derived the similarity transformation to approximate the solutions along with numerical results, whereas \citet{halpern1992dynamics} chose soluble surfactant-laden flow over a rigid bottom and ignored the sorption kinetics. As they considered a thin film, the vertical diffusion is very fast, and there is an uninterrupted flux from the free surface to the bottom of the film. They observed that the flux of the liquid layer depends on the film thickness, and the surface tension gradients decline sporadically, so over an extended period of time, an area of weak reverse flow emerges. \citet{jensen1993spreading} chooses the same model with consideration of sorption kinetics along with advection and diffusion to regulate the surfactant distribution. They found that the transient desorption of the surfactant from the free surface to the bulk flow slows down the spreading rate, but once the concentrations of the surfactant species are in equilibrium, a steep shock emerges in the film thickness upstream of the leading edge of the monolayer. \citet{jensen1994transport} extended their previous work \cite{jensen1993spreading} by allowing the bottom to absorb the solute and the transportation of the passive solute. They reported that transportation of solute by spreading a strip of insoluble surfactant-driven flow takes place in two stages, with the first part comprising vertical diffusion, and if at the end of this stage, solute is not absorbed by the bottom layer, then horizontal diffusion dominates in the second step. Moreover, the spreading of an insoluble surfactant monolayer as a planar strip and axisymmetric drop on deep viscous fluid is explored by \citet{jensen1995spreading}. They noticed that at the initial stage, a laminar subsurface boundary layer flow developed, and capillarity, gravity, and diffusion were influential at this stage, but at a sufficiently large prescribed time, the flow transitioned to a viscous-dominated flow. 
	
	Further, fingering instability identified in experimental studies led \citeauthor{matar1997linear} to a series \cite{matar1997linear, matar1998growth, matar1999development, matar1999spreading} of investigations on surfactant spreading over a thin film in the presence of external factors such as capillarity, diffusion, van der Waals forces, and Marangoni forces through linear stability and transient growth analyses along with direct numerical simulation for nonlinear equations. They observed that substantial growth can be achieved for significant van der Waals forces; however, large transient growth can be noticed with decay to follow if those forces are absent. \citet{fischer2003growth} chooses a different methodology to describe this same model and observed that large transient growth occurs if there is sufficient time for surfactant gradient and film thickness to develop their characteristic kink and shock at the leading edge of the monolayer. Also, these structures are sensitive to disturbances and usually appear in the spreading region. 
	
	The dynamics of the surfactant-laden flow are examined by \citet{edmonstone2004flow}, where they noted the formation of capillary ridges due to the Marangoni forces, and their transient growth analysis reveals that the film is linearly unstable to transverse disturbances with maximum growth at the moderate wavenumbers.  \citet{warner2004fingering} investigated the linear and nonlinear stability analysis of insoluble surfactant spreading over a thin film using transient growth analysis and direct numerical analysis. They observed prominent ridge formation at the leading edge, but due to mass conservation, the interface goes through thinning upstream of the leading edge. Moreover, they concluded that this thinning zone is unstable in the spanwise direction because of the fingering instability. The spreading of a surfactant-laden drop over a thin liquid layer is examined by \citet{jensen2006spreading}. They observed that the initial conditions have a major influence on the spreading rates, as the initial condition influences the contact line before the droplet. Moreover, thinning can appear in the neighbourhood of the contact line because the large surfactant gradient at the beginning can prevent the transport of the solute before the droplet.
	
	\citet{kovalchuk2016effect} performed an experimental study with two different types of surfactants, triethylene glycol monodecyl ether $C_{10} E O_3$ and trisiloxane superspreader, $(BT-278)$ to observe the wetting and adsorption kinetics. They reported that $BT-278$ drops spread more rapidly than the $C_{10} E O_3$ and mentioned that the flow due to the surface tension gradient induced Marangoni stress is a pivotal element in the spreading of the surfactant drops. Furthermore, the effect of the fluid thickness on surfactant spreading is explored by \citet{bergemann2018viscous}. They concluded that if the film thickness is very small, then the drop creates a steep front, which creates minimal disturbance to the fluid layer in front of it. In case of moderate thickness, the drop generates a wedge-like structure and produces substantial perturbation to the fluid layer ahead of the drop, and for a thick layer, the drop sinks into the layer and causes upwelling in the fluid layer and forms a transient bulge, and when this bulge is influenced by the gravity, the leading edge of the drop displays a wedge-like profile with a smaller angle.
	
	If the surface tension gradients caused by surfactant distribution are sufficiently large, then they can lead to the deformation of the thin film in the draining part to cause rupture \cite{jensen1992insoluble, braun2012dynamics, vivek2024rupture}. The study \cite{oron1997long} of stability of film rupture has been the subject of extensive research over the past few decades because of its applicability as a medium to transport aerosol in alveoli \cite{hermans2015lung}, assisting in cell adhesion in aqueous surroundings \cite{gallez1994non}, treatment of dry eye syndrome \cite{zhang2003analysis, braun2012dynamics} etc.  \citet{derjaguin1955definition} was one of the earliest works available in the literature. Then, extensive investigations \cite{de1994nonlinear, matar2002nonlinear, zhang2003analysis, zhang2003surfactant, dey2019model, yang2024rupture, yang2025vertical} are conducted in the subsequent years, where predominantly long-range van der Waals forces are considered as the major mechanism behind the rupture. These kinds of forces are feasible for a fluid of thickness $100-1000$ \AA. Moreover, the aforementioned research articles took into account other forces such as surface tension, surfactant, evaporation, etc. In the current study, soluble surfactant-laden flow is considered. In that case, desorption and adsorption between the interfacial and bulk surfactant concentrations are regulated through the flux condition, and this condition helps to reduce the effect of Marangoni stress (\cite{lin2000nonlinear, yang2025vertical} and the references therein). Previously, \citet{matar2002nonlinear} examined the rupture dynamics of soluble surfactant-laden flow. They derived the one-dimensional nonlinear film thickness equation in the rapid diffusion limit and high-order long-wave theory. In the case of slow diffusion, a one-dimensional bulk concentration model was also approximated. Their numerical results showed that the van der Waals forces, sorption kinetics rates, and surfactant solubility advance the rupture of the film. Recently, the stability analysis of soluble surfactant-contaminated trilayer thin film to observe the rupture of the thin film between two droplets during coalescence was explored by \citet{yang2024rupture}. They observed that surfactant solubility reduces the effect of Marangoni stress by diminishing interfacial surfactant gradients for symmetric outer-layer thickness, and the impact of soluble surfactant depends on the thickness of the layers. Most of the aforementioned studies subtly overlook the vertical concentration gradients by introducing rapid diffusion, which allows the concentration of the bulk surfactant to be represented by the cross-sectional average value. \citet{yang2025vertical} addresses this issue by considering a 2-dimensional description of the bulk surfactant concentration along with a traditional vertical averaging approximation and performs a comparative analysis between the two approaches. At initial stages of surfactant equilibrium, they noticed that the vertical average approximation inaccuracy for prediction of the rupture time is below $10$ \% for surfactant with $Pe \sim 10^4$, but that increased up to $40$ \% for surfactant with $Pe \sim$ with $10^6$.
	
	Most of the above studies focused on the spreading and rupture of thin film flow over solid substrates except \citet{zhang2003surfactant}. However, their study considered insoluble surfactant-laden flow. But, in general, the bottom surface is not smooth and can be very different from being solid. There are several studies that suggest the conventional model is not suited for hydrophilic liquid flow over hydrophobic boundaries at both nano- and micro-scales \cite{ruckenstein2018no, tretheway2002apparent, byun2008direct, tsai2009quantifying}. Experimental results of hydrophilic flow over hydrophobic surfaces are consistent with the Navier slip condition \cite{oron1997long, samanta2011falling}. Using the condition, \citet{bhat2018linear} investigated the stability analysis of insoluble surfactant-contaminated fluid flow over a slippery bottom. They observed different modes and noted that the P\'eclet number associated with the surfactant segregates the shear mode region into two different parts with contrasting behaviour. 
	
	On the other hand, externally imposed shear stress is one of the important mechanisms to regulate the thin film flow. \citet{smith1990mechanism} being one of the earliest to use external shear on the free surface in thin film flow dynamics. They reported that the stress induced by external shear influences the base flow and subsequently intensifies longitudinal flow disturbances. Due to its physical significance, such as bolus dispersal in the lungs during surfactant replacement therapy \cite{espinosa1999bolus} or airway closure method \cite{halpern1993surfactant}, it is used in numerous studies \cite{choudhury2024thermocapillary, hossain2025stability}. \citet{samanta2014shear} considered externally imposed shear as tangential stress destabilizes the flow near the criticality but stabilizes for large values of the Reynolds number. The influence of externally imposed shear on insoluble surfactant-coated thin film flow is investigated by \citet{bhat2019linear}. They observed three modes: surface mode, surfactant mode, and shear mode instability with external shear force, unstable surface, and surfactant modes. \citet{paul2025stability} investigates the effect of external shear for soluble surfactant-contaminated thin film flow over an inclined slippery plane. They recognized the three modes observed previously\cite{bhat2019linear, choudhury2024thermocapillary}. Moreover, they found out that the surfactant parameter $\beta_a$, and sorption kinetics parameter $k_a$ stabilize the surfactant mode, whereas the solubility parameter $R_a$ destabilizes the surfactant mode. 
	
	The above discussion indicates that the spreading phenomenon of surfactant-laden thin film flow is studied in detail, but the dual effect of the external shear force and slippery bottom on the spreading and rupture of soluble surfactant-laden flow has not been investigated till now, as per the knowledge of the authors. This motivates the authors to extend the work of \citet{jensen1993spreading} for externally imposed shear force and slippery bottom. In that context, the lubrication approximation is applied to the governing equations. Then, the evolution equations of the film thickness and surfactant concentration are derived and solved numerically. Section \ref{sec:Theory} is dedicated to the mathematical modeling, and the numerical investigation is enclosed in Section \ref{sec:NSD}. Then, the rupture mechanism is discussed in section \ref{sec:rup}. Finally, the summary and conclusion are given in Section \ref{sec:SC}.

	\section{MATHEMATICAL MODEL \& NON-DIMENSIONALISATION}
	\label{sec:Theory}
	An incompressible, Newtonian fluid filled with non-volatile soluble surfactant and flowing down a hydrophobic incline is considered here. An externally imposed shear force is also assumed to act along the free surface. The angle of inclination of the incline with the is $\xi$. The concentration of the bulk surfactant is assumed to be below CMC, so the emergence of a micelle is ignored. Also, the surfactant may dissolve in the bulk or be adsorbed at the surface. The dimensional governing equations along with the boundary conditions can be found in \citet{paul2025stability}. The characteristic length $\mathfrak{L}$ of the thin film flow is substantially greater than the depth $\mathfrak{H}$ of the fluid layer. So, 
	\bea
	\nonumber \varepsilon = \displaystyle \frac{\mathfrak{H}}{\mathfrak{L}} \ll 1
	\eea
	which allows the application of the lubrication approximation. 
	Now, before transforming the equations into dimensionless form under the framework of lubrication approximation, it is important to identify the dominant forces driving the system and create appropriate scales for the physical relevance of the scales.
	
	\begin{figure}[htbp!]
		\renewcommand{\figurename}{\scriptsize FIG.}
		\centering
		\includegraphics[height=7.0cm, width=18cm]{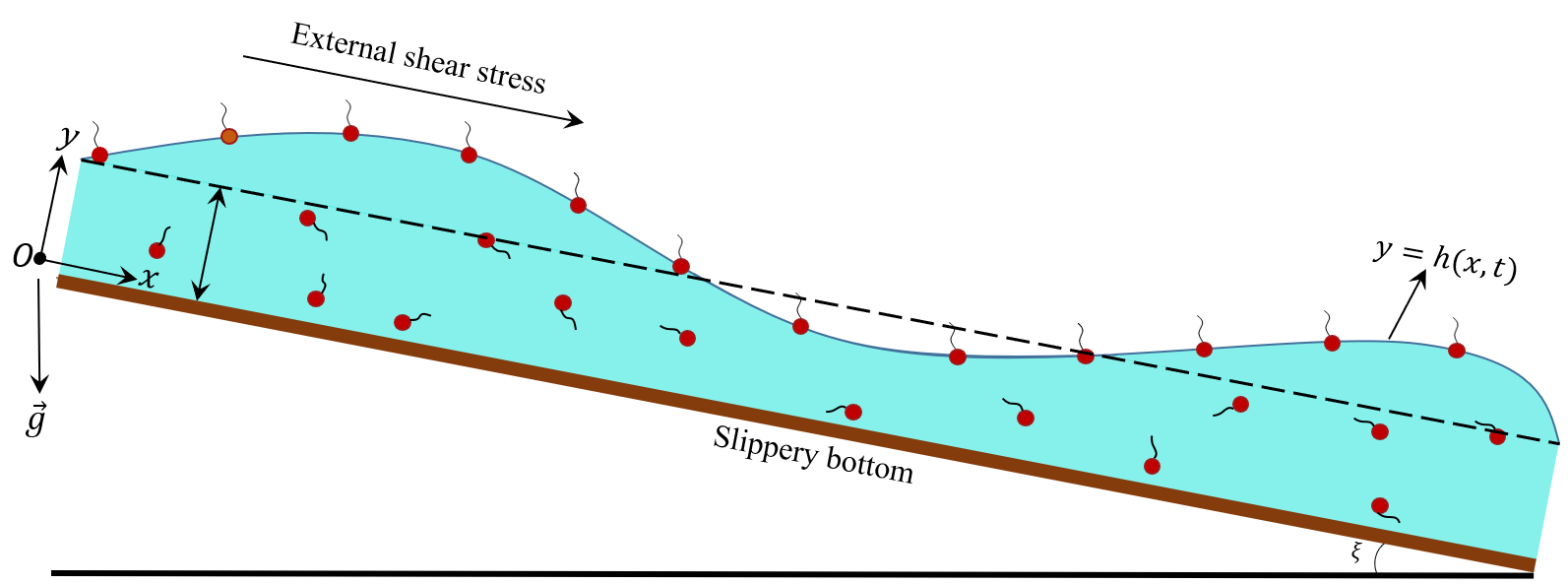}
		\caption{\scriptsize The schematic diagram of thin film fluid flow contaminated with soluble surfactant over a slippery inclined plane in the presence of external shear.}
		\label{fig:1}
	\end{figure}
	In the case of thin film flow, the van der Waals force is one of the major factors that instigate the disturbance. This kind of instability can be explained through height-dependent disjoining pressure. The external disturbances experienced by a sufficiently thin film (of thickness  $100-1000$~\AA~ as attractive van der Waals forces are operative for fluid thickness of that range) promote higher disjoining pressure at the troughs compared to the crests of the perturbed thin flow, which drives the fluid away from the troughs and intensifies the disturbances. However, viscosity and capillarity of the free surface promote stability by opposing the effect of van der Waals forces. The competition between these forces creates an imbalance of forces within the system, and after a threshold wavenumber, the dispersion forces become dominant, which leads to the rupture of the thin film \cite{de1994nonlinear, matar2002nonlinear, ruckenstein2018spontaneous, dey2019model, vivek2024rupture, yang2024rupture, yang2025vertical}. Also, it is feasible to estimate the wavelength at which the perturbation growth rate would attain the maxima that trigger the rupture \cite{matar2002nonlinear, dey2019model, vivek2024rupture, yang2024rupture, yang2025vertical}.

	Here, the planar distribution of the surfactant is considered for simplicity. Moreover, the difference in surface tension between the surfactant-free interface and the surfactant-adsorbed interface is $\mathfrak{S}$, which is the spreading coefficient. So, 
	\bea
	\nonumber \mathfrak{S} = \sigma_c - \sigma_m
	\eea
	where $\sigma_c$ and $\sigma_m$ represent the surface tension of the clean fluid and surfactant-absorbed interface, respectively. Then the following non-dimensional scales: 
	\bea
	\nonumber x = \mathfrak{L} \bar{x};~~ y = \mathfrak{H} \bar{y};~~h = \mathfrak{H} \bar{h};~~ u = \mathfrak{U} \bar{u};~~ v = \varepsilon \mathfrak{U} \bar{v};~~ t = \frac{\mathfrak{L}}{\mathfrak{U}} \bar{t};~~ p = \displaystyle \frac{\mu_{visc} \mathfrak{U} \mathfrak{L}}{\mathfrak{H}^2} \bar{p};\\
	\nonumber \Gamma_a = \Gamma_{a, eq} \bar{\Gamma}_a;~~\Gamma = \displaystyle \frac{k_1}{k_2 \Gamma_{a, eq}} \bar{\Gamma};~~ J_{ba} = \frac{\mathfrak{U} \Gamma_{a, eq}}{\mathfrak{L}} \bar{J}_{ba};~~ \mu = \frac{\mu_{visc} \mathfrak{U}}{\mathfrak{H}} \bar{\mu};~~\sigma = \sigma_m + \mathfrak{S} \bar{\sigma };\\
	\nonumber \mathfrak{U} = \displaystyle \frac{\mathfrak{S} \mathfrak{H}}{\mu_{visc} \mathfrak{L}};~~ \Sigma = \frac{\sigma_c}{\sigma_m};~\delta = \frac{\delta_s}{\mathfrak{H}};k_a = \frac{k_2 \mathfrak{L}}{\mathfrak{U}};
	\eea
	are used for the applicability of lubrication approximation theory. Here, $k_1$ is the adsorption rate and $k_2$ is the desorption rate. It is noteworthy to mention the potential energy $\varphi $ per volume in the liquid due to the effect of the van der Waals force. The potential energy is defined \cite{jensen1992insoluble, zhang2003analysis, ruckenstein2018spontaneous} as
	\bea
	\nonumber \varphi(h) = \displaystyle \frac{\mathcal{A}}{6 \pi h^3}
	\eea
	with $\mathcal{A}$ the Hamaker constant, and its dimensionless form is defined by $\mathcal{A} = 6 \pi \mathfrak{S} \mathfrak{H}^2 \bar{\mathcal{A}}$.
	
	The dimensionless Shelduko equation of state \cite{sheludko1967thin, jensen1992insoluble, paul2025stability} is considered 
	\bea
	\nonumber \sigma = (\alpha+1) \left[ 1 + \theta(\alpha) \Gamma_a \right]^{-3} - \alpha
	\eea
	for surface tension profile. Here, $\theta(\alpha) = \displaystyle \left(\frac{\alpha +1}{\alpha}\right)^{1/3}-1$ with $\alpha = \displaystyle \frac{\sigma_m}{\mathfrak{S}}$.\\
	Using the above scaling, all the governing equations and the boundary conditions are made dimensionless, and following the lubrication approximation \cite{jensen1992insoluble, yang2025vertical}, the terms of $\mathcal{O}(\varepsilon^2)$ are ignored for brevity. Moreover, throughout the article, the distribution of surfactant is considered to be of a planar type. Also, the overhead bars are suppressed from the dimensionless variables. The free surface or interface is considered at $y = h(x,t)$. Hence, dimensionless governing equations at the leading order $(\mathcal{O}(1))$ are
	\bea
	u_x + v_y = 0, \label{eqn1}\\
	u_{yy} + Bo = (p + \varphi)_x, \label{eqn2}\\
	p_y = 0, \label{eqn3}
	\eea
	where $Bo = \displaystyle \frac{\rho g \sin(\xi) \mathfrak{H} \mathfrak{L}}{\mathfrak{S}}$.
	The dimensionless tangential stress condition at $y = h(x,t)$ will be
	\bea
	u_y = \sigma_x + \mu. \label{eqn4}
	\eea
	and the kinematic boundary condition is 
	\bea
	v = h_t + u h_x, \label{eqn5}
	\eea
	The leading-order dimensionless normal stress boundary condition is
	\bea
	p\bigg|_{y=h} = -\mathcal{C} h_{xx}, \label{eqn6}
	\eea
	where $\mathcal{C} = \varepsilon^2 Ca = \varepsilon^2 \displaystyle \frac{\sigma_m}{\mathfrak{S}}$ corresponds to the capillary effect.\\
	The dimensionless interface surfactant transport equation at the interface $\mathcal{O}(1)$ is
	\bea
	\Gamma_{a,t} + \left( u \Gamma_a \right)_x = \displaystyle \frac{1}{Pe_s} \Gamma_{a, xx} + J_{ba} \label{eqn7}
	\eea
	and the bulk surfactant transportation equation is
	\bea
	\Gamma_t + u \Gamma_x + v \Gamma_y = \displaystyle \frac{1}{Pe_b} \left( \Gamma_{xx} + \frac{1}{\varepsilon^2} \Gamma_{yy} \right). \label{eqn8}
	\eea
	Here, $Pe_s = \displaystyle \frac{\mathfrak{U} \mathfrak{L}}{D_s}$ and $Pe_b = \displaystyle \frac{\mathfrak{U} \mathfrak{L}}{D_b}$ are the P\'eclet numbers corresponding to the interface and the monomers in the bulk, respectively, with $D_s$, and $D_b$ being the diffusivity coefficients corresponding to the interface and bulk surfactant, respectively. It is to be noted that \eqref{eqn8} retains a term $\mathcal{O}(1/\varepsilon^2)$, which is kept intentionally, as the rapid vertical diffusion will be explored in Section \ref{sec:CSA} where $\varepsilon^2 Pe_b \ll 1$ is considered. \\
	The Navier-slip \cite{samanta2011falling} and no penetration boundary conditions at the bottom are
	\bea
	u = \delta u_y;~~v = 0 \label{eqn9}
	\eea
	At the beginning of the surfactant deposition, it is assumed that it behaves as a surface monolayer. Also, during the transitional phase when surfactant molecules desorb from the interface and enter the bulk fluid, linearized sorption kinetics are considered. Thus, the flux of surfactant at the interface can be described as 
	\bea
	J(\Gamma_a, \Gamma\bigg|_{y=h}) = -\frac{1}{\beta_a Pe_b} (\textbf{n}\cdot \grad \Gamma)\bigg|_{y = h} = k_a \left\{\Gamma\bigg|_{y=h} \left( 1 - \Gamma_a \right) - \Gamma_a \right\}~~\label{eqn10}
	\eea
	where $\Gamma_a$, $\Gamma\bigg|_{y=h}$ are the dimensionless surfactant concentrations of the interface and bulk. Now, if $k_a = k_2 T$ is the ratio between the time scale of the flow and the time scale of desorption, then after a period of $\mathcal{O}(1/k_a)$, the interface and bulk surfactant concentrations reach an instantaneous equilibrium due to the linearized Langmuir isotherm and the adsorption and desorption fluxes balance. That gives $\Gamma_a = (k_1/k_2) \Gamma\bigg|_{y=h}$ in dimensional form and becomes  $\Gamma_a = \Gamma\bigg|_{y=h}$ in dimensionless form. 
	The flux equation \eqref{eqn10} can be written as 
	\bea
	J_{ba} = \displaystyle \frac{1}{\beta_a} \left(- \frac{1}{\varepsilon^2 Pe_b} \Gamma_y + \frac{1}{Pe_b} h_x \Gamma_x \right)\bigg|_{y=h} = k_a \left\{\Gamma\bigg|_{y=h} \left( 1 - \Gamma_a \right) - \Gamma_a \right\}, \label{eqn11}
	\eea
	where $\beta_a = \displaystyle \frac{\Gamma_{a \infty}}{\mathfrak{H} \Gamma_{CMC}} = \frac{k_1}{k_2 \mathfrak{H}}$ is the ratio between the rates of adsorption and desorption. That means if $\beta_a$ is large, then the adsorption rate is high, and the surfactant concentrates near the interface and behaves as an insoluble surfactant. However, if $\beta_a$ is small, then the desorption rate dominates the adsorption rate, and the surfactant dissolves within the bulk fluid. Finally, if there is no flux of surfactant at the bottom, then
	\bea
	\Gamma_y = 0, ~~\text{at}~~ y = 0. \label{eqn12}
	\eea 
	The no flux condition at the bottom guarantees the conservation of the total mass of the different species of surfactant from the equations \eqref{eqn7} and \eqref{eqn8}, so
	\bea
	\mathcal{M} = \int_{0}^{\infty} \Gamma_a dx + \frac{1}{\beta_a} \int_{0}^{\infty} \left(\int_{0}^{h(x,t)} \Gamma dz \right) dx
	\eea
	\subsection{Evolution equation}
	This subsection is devoted to deriving the evolution equation of the film thickness. First of all, the equation \eqref{eqn3} is solved using the appropriate boundary conditions at \eqref{eqn6} which gives
	\bea
	p(x,y,t) = - \mathcal{C} h_{xx}
	\eea
	Then, solving the equation \eqref{eqn2} with boundary conditions at \eqref{eqn4} \& \eqref{eqn9}, we get
	\bea
	u(x,y,t) = \left[ -\frac{3 (\alpha+1) \theta(\alpha)}{(1+ \theta(\alpha) \Gamma_a)^4} \Gamma_{a,x} + \mu \right] (y + \delta) + \left\{ Bo + \left( \mathcal{C} h_{xx} - \frac{\mathcal{A}}{h^3} \right)_x \right\} \left\{ h \delta + (h-\delta) y - \frac{y^2}{2} \right\}. \label{eqn15}
	\eea
	
	Now, substituting the expression \eqref{eqn15} on the kinematic equation \eqref{eqn5},  
	\bea
	h_t + \displaystyle\left[ \left\{ -\frac{3 (\alpha+1) \theta(\alpha)}{(1+ \theta(\alpha) \Gamma_a)^4} \Gamma_{a,x} + \mu \right\} \left( \frac{h^2}{2} + \delta h \right) + \left\{ Bo + \left( \mathcal{C} h_{xx} - \frac{\mathcal{A}}{h^3} \right)_x \right\} \left( \frac{1}{3} h^3 + \frac{\delta}{2} h^2 \right) \right]_x = 0, \label{eqn16}
	\eea
	and the evolution equation \eqref{eqn7} for surfactant monomers,
	\bea
	& \nonumber \Gamma_{a,t} + \displaystyle \left[ \left\{ \left( -\frac{3 (\alpha+1) \theta(\alpha)}{(1+ \theta(\alpha) \Gamma_a)^4} \Gamma_{a,x} + \mu\right) (h + \delta) + \left( Bo + \left( \mathcal{C} h_{xx} - \frac{\mathcal{A}}{h^3} \right)_x \right) \frac{h^2}{2} \right\} \Gamma_a \right]_x = \displaystyle \frac{1}{Pe_s} \Gamma_{a, xx}\\
	& + k_a \left[  \Gamma\bigg|_{y=h} \left( 1 - \Gamma_a \right) - \Gamma_a \right] \label{eqn17}
	\eea
	\subsection{Cross-Sectional Average} \label{sec:CSA}
	The spreading of aqueous surfactant solution on thin film flow accompanies the risk of the emergence of fingering instability. This phenomenon can be smoothed out by eradicating vertical variations. To do that, rapid vertical diffusion can be a useful tool. For that $\varepsilon^2 Pe_b \ll 1$ is considered. Then, the surfactant bulk concentration can be segregated into components comprising $\Gamma_0(x, t)$ independent of $y$ and an infinitesimal fluctuation $\Gamma_1(x, y, t)$ that has zero vertical average, such as
	\bea
	\Gamma(x, y, t) = \Gamma_0(x, t) + \varepsilon^2 \Gamma_1(x, y, t), \label{eqn18}
	\eea
	with 
	\bea
	\bar{\Gamma}_1 = \displaystyle \frac{1}{h} \int_{0}^{h} \Gamma_1(x, y, t) dy = 0 \label{eqn19}
	\eea
	as considered in the literature\cite{warner2004fingering, jensen1993spreading, mavromoustaki2012dynamics}. Substituting \eqref{eqn18} in \eqref{eqn7} results in,
	\bea
	& \nonumber \Gamma_{a,t} + \left[ \left\{ \left\{ \left( -\frac{3 (\alpha+1) \theta(\alpha)}{(1+ \theta(\alpha) \Gamma_a)^4} \Gamma_{a,x} + \mu \right) (h + \delta) + \left( Bo + \left( \mathcal{C} h_{xx} - \frac{\mathcal{A}}{h^3} \right)_x \right) \frac{h^2}{2} \right\} \right\} \Gamma_a \right]_x \\
	& = \displaystyle \frac{1}{Pe_s} \Gamma_{a, xx} + k_a \left[ \Gamma_0 (1 - \Gamma_a) - \Gamma_a \right] + \mathcal{O}(\varepsilon^2). \label{eqn20}
	\eea
	Again, the expression \eqref{eqn18} is substituted in the equation \eqref{eqn8} and then integrated with respect to `$y$' over the interval $[0, h]$ and then divided by the film thickness gives,
	\bea
	& \nonumber \Gamma_{0,t} + \left[ \left\{ -\frac{3 (\alpha+1) \theta(\alpha)}{(1+ \theta(\alpha) \Gamma_a)^4} \Gamma_{a,x} + \mu \right\} \left( \frac{h}{2} + \delta  \right) + \left\{ Bo + \left( \mathcal{C} h_{xx} - \frac{\mathcal{A}}{h^3} \right)_x \right\} \left( \frac{1}{3} h^2 + \frac{\delta}{2} h \right) \right] \Gamma_{0,x}\\
	& - \displaystyle \frac{(h \Gamma_{0,x})_x}{h Pe_b} + \frac{\beta_a}{h} k_a \left[  \left( 1 - \Gamma_a \right) \Gamma_0 - \Gamma_a \right] = \mathcal{O}(\varepsilon^2) \label{eqn21}
	\eea
	Now, for simplicity, all the $\mathcal{O}(\varepsilon^2)$ terms are ignored. Hence, the evolution equations of film thickness and surfactant concentration of the interface and the bulk are given by \eqref{eqn16}, \eqref{eqn20}, \eqref{eqn21}, respectively.
	If $k_a \rightarrow 0$, then there will be no sorption kinetics, and the problem reduces to insoluble surfactant-laden thin film flow. However, if $k_a \rightarrow \infty$, then sorption rates are extremely fast, and the bulk and surface surfactants reach equilibrium such that $\Gamma_a(x,t) = \Gamma_0(x,t)$ at the leading order. But still, there exists a brisk diffusive flux between the surface and bulk surfactant concentrations, which acts on any spatial or temporal disturbances. So, eliminating the flux terms between the equations \eqref{eqn19} \& \eqref{eqn20},
	\bea
	& \nonumber h \left\{ \Gamma_{0,t} + \left[ \left\{ -\frac{3 (\alpha+1) \theta(\alpha)}{(1+ \theta(\alpha) \Gamma_a)^4} \Gamma_{a,x} + \mu \right\} \left( \frac{h}{2} + \delta  \right) + \left\{ Bo + \left( \mathcal{C} h_{xx} - \frac{\mathcal{A}}{h^3} \right)_x \right\} \left( \frac{1}{3} h^2 + \frac{\delta}{2} h \right) \right] \Gamma_{0,x} - \displaystyle \frac{(h \Gamma_{0,x})_x}{h Pe_b} \right\}\\
	& + \beta_a \left\{ \Gamma_{a,t} + \left[ \left\{ \left\{ \left( -\frac{3 (\alpha+1) \theta(\alpha)}{(1+ \theta(\alpha) \Gamma_a)^4} \Gamma_{a,x} + \mu\right) (h + \delta) + \left( Bo + \left( \mathcal{C} h_{xx} - \frac{\mathcal{A}}{h^3} \right)_x \right) \frac{h^2}{2} \right\} \right\} \Gamma_a \right]_x - \displaystyle \frac{1}{Pe_s} \Gamma_{a, xx} \right\} = 0, \label{eqn22}
	\eea
	Now, if the stream function is defined as
	\bea
	u = \displaystyle \frac{\partial \psi}{\partial y};~~v = - \frac{\partial \psi}{\partial x}
	\eea
	then, the stream function can be calculated as
	\bea
	\psi(x,y,t) = \int_{0}^{y} u(x,y,t) dy.
	\eea
	\section{Numerical simulations and discussion} \label{sec:NSD}
	The set of equations \eqref{eqn16}, \eqref{eqn20},\eqref{eqn21} reveals that there will be an evanescent period of $\mathcal{O}(1/k_a)$ when the surfactant species will come to local equilibrium. Thus, the equations \eqref{eqn16}, \eqref{eqn22} can be used to describe the dynamics of the system. Hence, the spreading of the surfactant-laden flow will be analyzed in two perspectives: $(i)$ for rapid sorption kinetics $k_a \gg 1$, $(ii)$ for slow sorption kinetics $k_a \ll 1$. Moreover, the distribution of surfactant is one of the important aspects that needs to be considered. Here, two types, (i) strip and (ii) front distributions, are taken into account following \citet{jensen1992insoluble, jensen1993spreading}. In each scenario, the equations are solved numerically using a central difference scheme for the spatial variable and Gear's method \cite{gaver1990dynamics, jensen1993spreading} in time.
	\subsection{Rapid sorption kinetics $(k_a \gg 1)$}
	
	The impact of gravity, van der Waals forces, and capillarity is ignored during the exploration of the spreading phenomena, so the Marangoni forces and viscous effect will be two possible forces impacting the spreading of surfactant in the case of rapid sorption kinetics. One of the distributions is a planar strip for which the no-flux boundary condition is imposed at $x = 0$, and the other one is an advancing front with a flux boundary condition applied at $x = 0$. To gain sufficient numerical advantages, following \citet{jensen1993spreading}, the evolution equations are transformed.
	
	\subsubsection{Strip}
	In case of $\beta_a \rightarrow \infty$ and for large values of the interface surfactant P\'eclet number $Pe_s$, the width of the spreading surfactant monolayer strip grows proportionally to $t^{1/3}$. This scaling is even valid for soluble surfactant-contaminated flow \cite{jensen1993spreading}. So, define,
	\bea
	\zeta = \displaystyle \frac{x}{t^{1/3}};~~\tau = t;~~\mathcal{H}(\zeta, \tau) = h(x,t);~~ \mathcal{F}(\zeta, \tau) = t^{1/3} \Gamma_0(x,t)
	\eea
	and the equations \eqref{eqn16} \& \eqref{eqn22} converts into
	\bea
	\displaystyle \tau \mathcal{H}_{\tau} - \frac{1}{3} \zeta \mathcal{H}_{\zeta} + \left[ \frac{1}{2} \left( - \frac{3 (\alpha) \theta(\alpha) \tau^{4/3}}{ (\tau^{1/3} + \theta(\alpha) \mathscr{F})^4} \mathcal{F}_{\zeta} + \mu \tau^{2/3} \right) \left( \mathcal{H}^2 + 2 \delta \mathcal{H} \right) \right]_{\zeta} = 0,\label{eqn26} \\
	\nonumber \displaystyle \mathcal{H} \left\{ \tau \mathcal{F}_{\tau} - \frac{1}{3}(\zeta \mathcal{F})_{\zeta} + \left[ \left( - \frac{3 (\alpha + 1) \theta(\alpha) }{2 \left(\tau^{1/3} + \theta(\alpha) \mathcal{F}\right)^4} \mathcal{F}^2_{\zeta} + \frac{\mu}{2} \tau^{1/3} \right) \left( \mathcal{H} + 2 \delta \right) \right] \mathcal{F}_{\zeta} - \frac{\tau^{1/3} \left( \mathcal{H} \mathcal{F}_{\zeta} \right)_{\zeta}}{\mathcal{H} Pe_b} \right\} \\
	+ \beta_a \left\{  \tau \mathcal{F}_{\tau} - \frac{1}{3}(\zeta \mathcal{F})_{\zeta} + \left[ \left( - \frac{3 (\alpha + 1) \theta(\alpha) }{ \left(\tau^{1/3} + \theta(\alpha) \mathcal{F}\right)^4} \mathcal{F}_{\zeta} + \mu \tau^{1/3} \right) (\mathcal{H} + \delta) \mathcal{F} \right]_{\zeta} + \frac{1}{Pe_s} \tau^{1/3} \mathcal{F}_{\zeta \zeta} \right\} = 0. \label{eqn27}
	\eea
	The equations \eqref{eqn26} \& \eqref{eqn27} are solved using the aforementioned method, and the initial condition used here is
	\bea
	\mathcal{H}(\zeta, 1) = 1;~~\mathcal{F}(\zeta, 1) = \displaystyle \frac{1}{2} \left[ 1 - \tanh(\frac{\zeta - 0.5}{0.1}) \right]. \label{eqn28}
	\eea
	In order to solve the above boundary value problem, Gear's method based on the built-in solver ``GMRES'' is used from the package  ``\hyperref{https://iterativesolvers.julialinearalgebra.org/stable/}{}{}{IterativeSolvers.jl}'' using the Julia programming language \cite{bezanson2017julia}.

	\begin{figure}[htbp!]
		\renewcommand{\figurename}{\scriptsize FIG.}
		\centering
		\includegraphics[height=6.0cm, width=8.5cm]{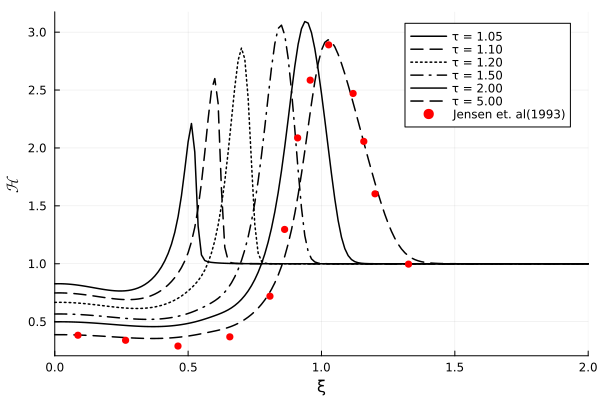}
		\includegraphics[height=6.0cm, width=8.5cm]{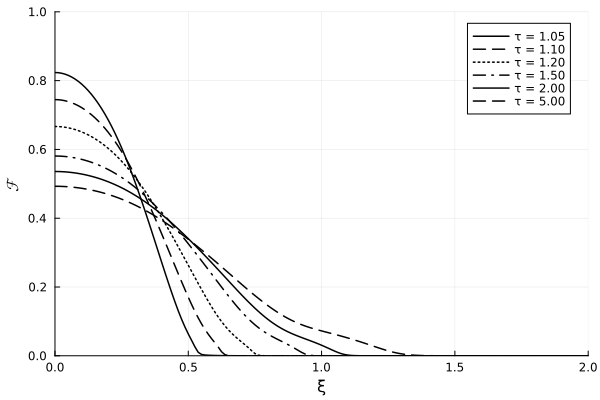}
		\caption{\scriptsize The (a) $\mathcal{H}$, and (b) $\mathcal{F}$ profiles for a soluble surfactant-laden spreading strip in case of fast sorption kinetics by solving the set of equations \eqref{eqn26}-\eqref{eqn27} subjected to the boundary conditions \eqref{eqn28}. These profiles are obtained when $\sigma_x = \Gamma_x$ as was considered by \citet{jensen1993spreading} in the absence of external shear force ($\mu = 0.0$) and no-slip bottom condition (i.e., $\delta = 0.0$). The profiles are plotted at time $\tau = 1.05,~1.1,~1.2,~1.5,~2.0,~5.0$. The red bullets are the data points obtained from Figure 2(a) of \citet{jensen1993spreading} for $\tau = 5.0$. The relevant parameters are $Pe_s = 100.0$, $Pe_b = 100.0$, $\xi_m = 0.25$, $\xi_w = 0.1$}
		\label{fig:2}
	\end{figure}
	
	The $\mathcal{H}$ and $\mathcal{F}$ profiles in the figures (\ref{fig:2}) are plotted in the absence of externally imposed shear force ($\mu = 0.0$) and slip length ($\delta = 0.0$), along with a linear interface surfactant profile ($ \sigma_x = - \Gamma_x$). The red bullets are data points obtained from Figure 2(a) of \citet{jensen1993spreading} for $\tau = 5.0$. It can be seen that the results are in good agreement. Figure \ref{fig:2}(a) shows that, as the planar strip of surfactant spreads, the film surges at the leading edge of the surfactant monolayer, and a thinning in the film thickness occurs behind the surge. A shock-like structure is formed initially but smoothed as time advances due to diffusion and capillarity. The surfactant concentration maintains a fairly uniform distribution except at the shock and at the upstream end.\\
	
	\begin{figure}[ht!]
		\begin{center} 
			\subfigure[$\delta = 0.00$]{\label{fig:3_a}\includegraphics*[height=4.5cm, width=5.5cm]{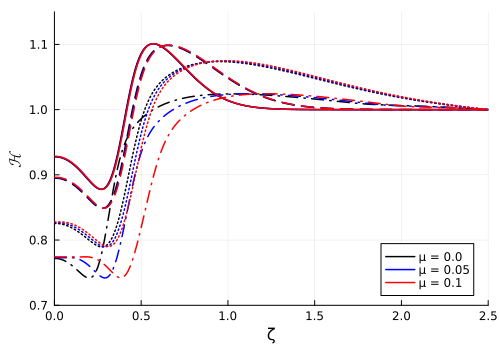}}
			\subfigure[$\delta = 0.04$]{\label{fig:3_b}\includegraphics*[height=4.5cm, width=5.5cm]{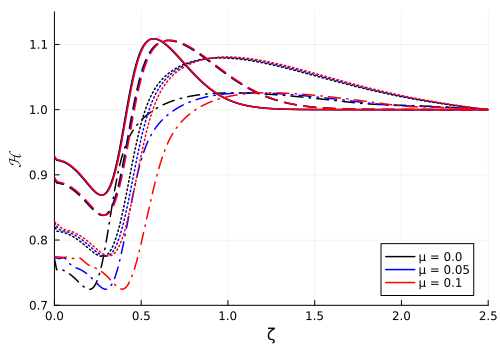}}
			\subfigure[$\delta = 0.08$]{\label{fig:3_c}\includegraphics*[height=4.5cm, width=5.5cm]{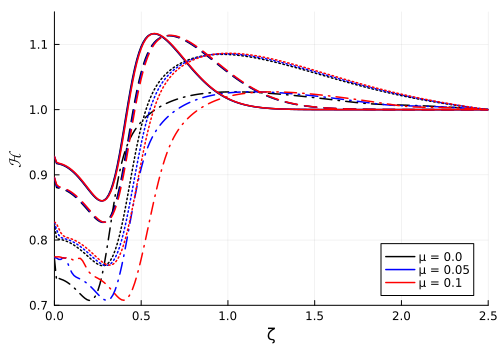}}
			\subfigure[$\delta = 0.00$]{\label{fig:3_d}\includegraphics*[height=4.5cm, width=5.5cm]{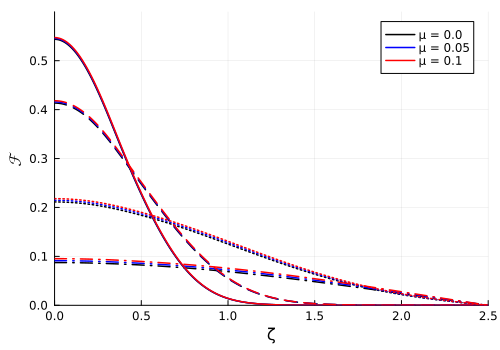}}
			\subfigure[$\delta = 0.04$]{\label{fig:3_e}\includegraphics*[height=4.5cm, width=5.5cm]{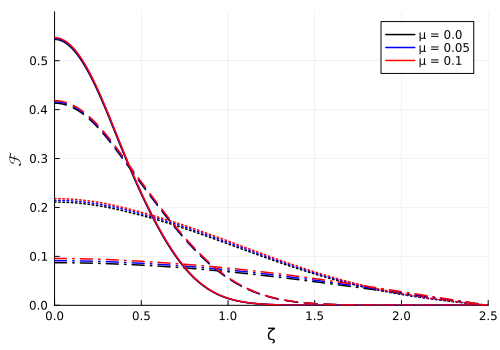}}
			\subfigure[$\delta = 0.08$]{\label{fig:3_f}\includegraphics*[height=4.5cm, width=5.5cm]{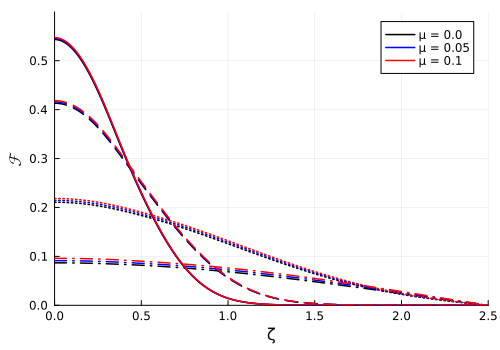}}
		\end{center} 
		\caption{ \scriptsize The variation of (a)-(c) $\mathcal{H}$, (d)-(f) $\mathcal{F}$ profile at $\delta = 0.00, 0.04, 0.08$ respectively for a soluble surfactant-laden spreading strip in the case of fast sorption kinetics by varying the externally imposed shear force. The solid, dashed, dotted, and dash-dot curves correspond to $\tau = 1.05,~ 1.1,~ 1.5,~ 5.0$. The relevant parameters are $\beta_a = 1.0$, $Pe_s = 1.0$, $Pe_b = 1.0$, $\alpha = 0.25$, $\xi_m = 0.25$, $\xi_w = 0.1$. } \label{fig:3}
	\end{figure}
	
	The figures (\ref{fig:3}) illustrate the effects of externally imposed shear $\mu$ and slip parameters $\delta$ on the spreading dynamics of a strip-type distribution of soluble surfactant thin-film flow. In the absence of a slippery bottom (see Figures (\ref{fig:3_a} - \ref{fig:3_c})), the film thickness experiences Marangoni thinning at the surfactant deposition zone, as a surfactant concentration gradient influences the liquid outward, which creates a capillary ridge at the leading edge of the monomer. However, with the introduction of a slippery bottom, the resistance at the bottom reduces, and the base of the capillary ridge undergoes less viscous drag. Hence, the fluid front spreads faster and advances horizontally sooner compared to the no-slip scenario. On the other hand, the introduction of external shear acts as an auxiliary driving force besides Marangoni stress. Consequently, the increased external shear amplifies the thinning of the film and drives the capillary ridge at a faster rate. As the slip parameter $\delta$ and external shear $\tau$ drive the spreading of the surfactant much faster, the state of equilibrium between the surfactant at the interface and the bulk due to the rapid sorption kinetics dilutes the localized concentration peaks over time (see figures (\ref{fig:3_d} - \ref{fig:3_f})).

	\begin{figure}[ht!]
		\begin{center} 
			\subfigure[$\mu = 0.0$]{\label{fig:4_a}\includegraphics*[height=4.5cm, width=5.5cm]{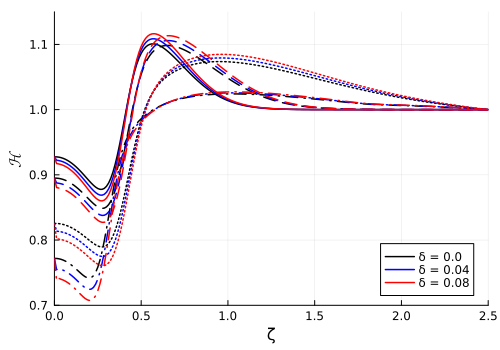}}
			\subfigure[$\mu = 0.05$]{\label{fig:4_b}\includegraphics*[height=4.5cm, width=5.5cm]{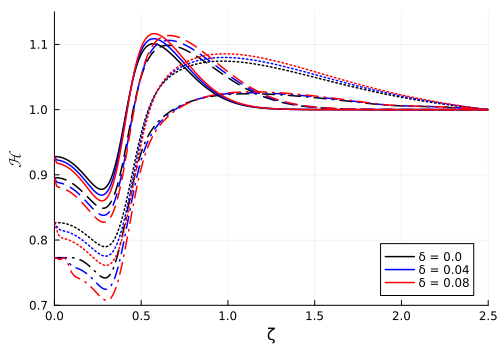}}
			\subfigure[$\mu = 0.1$]{\label{fig:4_c}\includegraphics*[height=4.5cm, width=5.5cm]{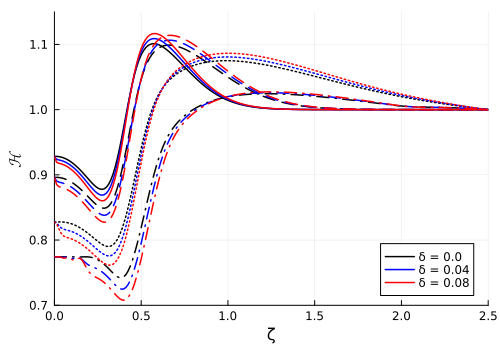}}
			\subfigure[$\mu = 0.0$]{\label{fig:4_d}\includegraphics*[height=4.5cm, width=5.5cm]{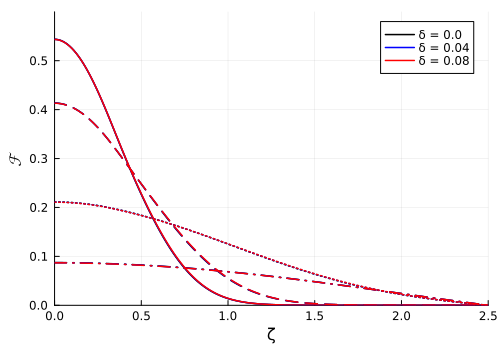}}
			\subfigure[$\mu = 0.05$]{\label{fig:4_e}\includegraphics*[height=4.5cm, width=5.5cm]{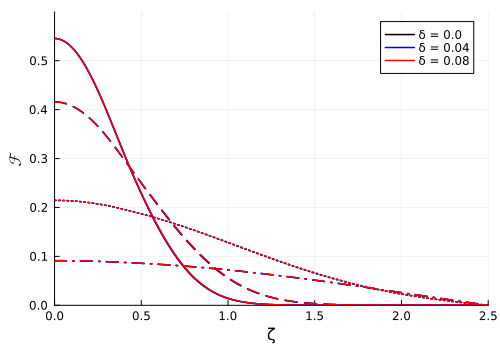}}
			\subfigure[$\mu = 0.1$]{\label{fig:4_f}\includegraphics*[height=4.5cm, width=5.5cm]{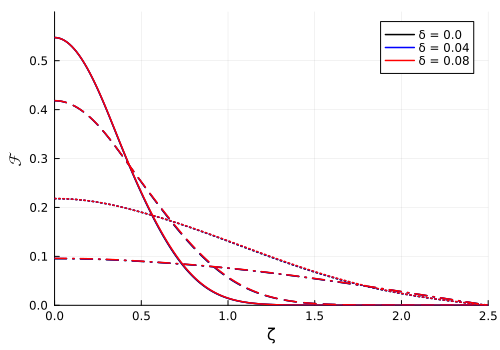}}
		\end{center} 
		\caption{ \scriptsize The variation of (a)-(c) $\mathcal{H}$, (d)-(f) $\mathcal{F}$ profile at $\mu = 0.0, 0.05, 0.1$ respectively for a soluble surfactant-laden spreading strip in the case of fast sorption kinetics by varying the slip length. The solid, dashed, dotted, and dash-dot curves correspond to $\tau = 1.05,~ 1.1,~ 1.5,~ 5.0$. The relevant parameters are $\beta_a = 1.0$, $Pe_s = 1.0$, $Pe_b = 1.0$, $\alpha = 0.25$, $\xi_m = 0.25$, $\xi_w = 0.1$. } \label{fig:4}
	\end{figure}
	
	In the following figures (\ref{fig:4}), the film thickness $\mathcal{H}$ and the corresponding surfactant concentration $\mathcal{F}$ are plotted for different external shear $\mu$. In the absence of a slippery bottom (Figures \ref{fig:4_a}-\ref{fig:4_c}), the higher viscous drag near the bottom reduces the amount of fluid that can be accumulated at the shock at the leading edge of the monomer deposition. But as soon as a hydrophobic substrate is considered, the viscous drag or the friction at the bottom reduces, a greater volume of fluid is transported, and it piles up in the shock front and heightens it. In contrast, in the absence of external shear force ($\mu = 0.0$), the shock is primarily driven by the localized surfactant gradient or Marangoni stress, and slip at the bottom increases the shock amplitude and flattens the film. However, the positive external shear force provides a constant momentum at the interface, and paired with the slip parameter, the film experiences less resistance at the bottom, which in turn shows an advancing front and sharper capillary ridge. The surfactant concentration $\mathcal{F}$ attains the peak value, and with increasing slip length, the peak drops more quickly, but the surfactant distribution spreads over a substantially wider spatial domain (see Figures \ref{fig:4_d}-\ref{fig:4_f}). The positive slip length increases the convective transport of the fluid interface, and since the surfactant is distributed in a faster interface, the distribution of the surfactant is spread much more quickly.

	\begin{figure}[ht!]
		\begin{center} 
			\subfigure[$\mu = 0.00, \delta = 0.00$]{\label{fig:5_a}\includegraphics*[height=4.5cm, width=5.5cm]{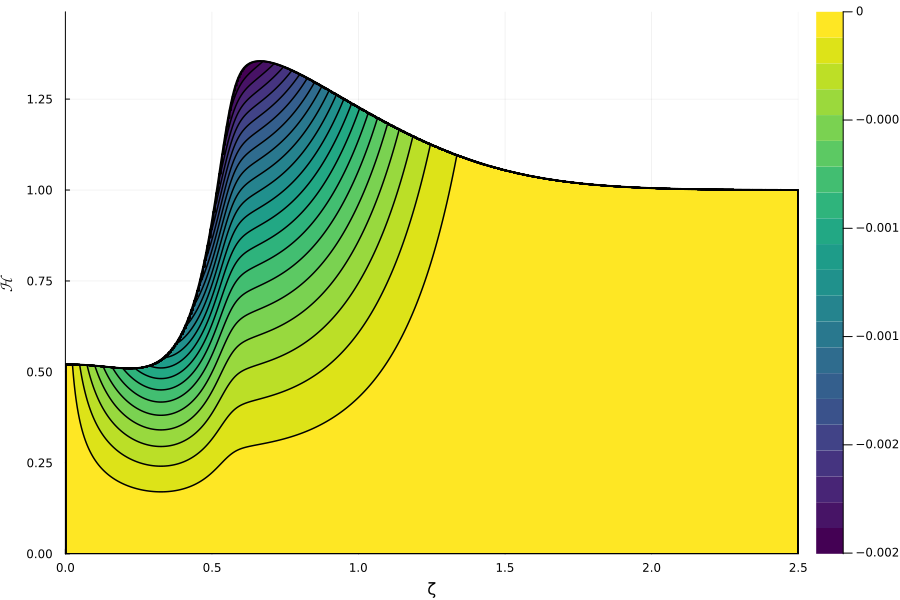}}
			\subfigure[$\mu = 0.00, \delta = 0.04$]{\label{fig:5_b}\includegraphics*[height=4.5cm, width=5.5cm]{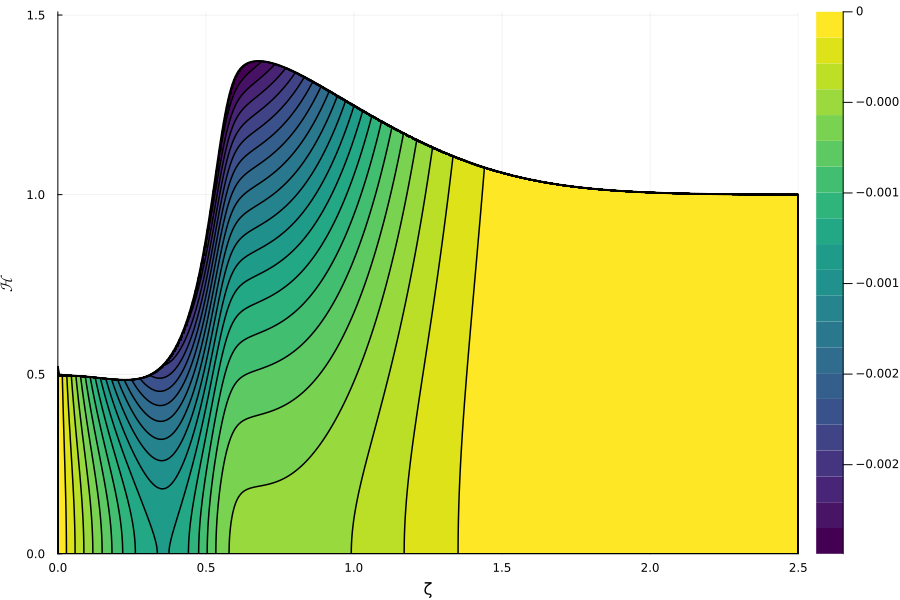}}
			\subfigure[$\mu = 0.00, \delta = 0.08$]{\label{fig:5_c}\includegraphics*[height=4.5cm, width=5.5cm]{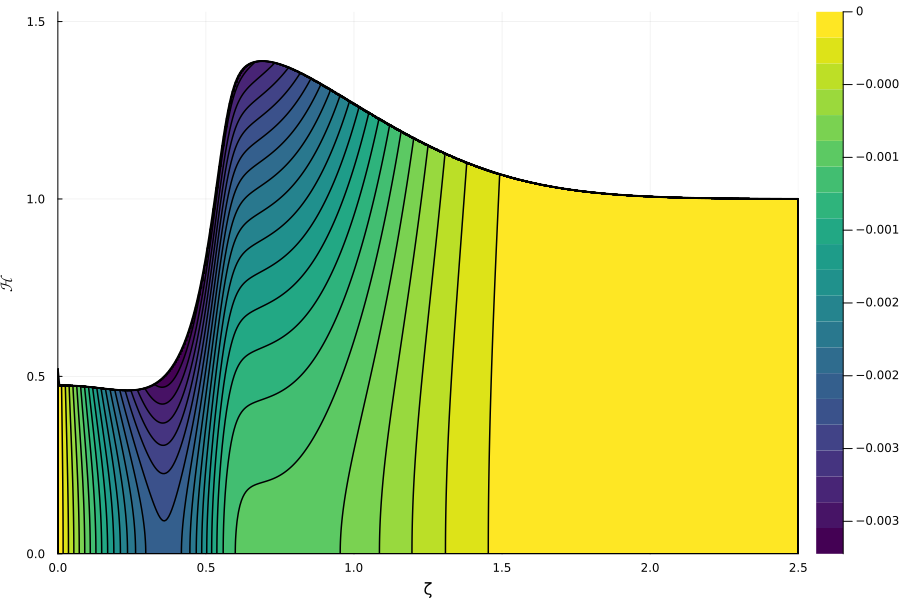}}
			\subfigure[$\mu = 0.05, \delta = 0.00$]{\label{fig:5_d}\includegraphics*[height=4.5cm, width=5.5cm]{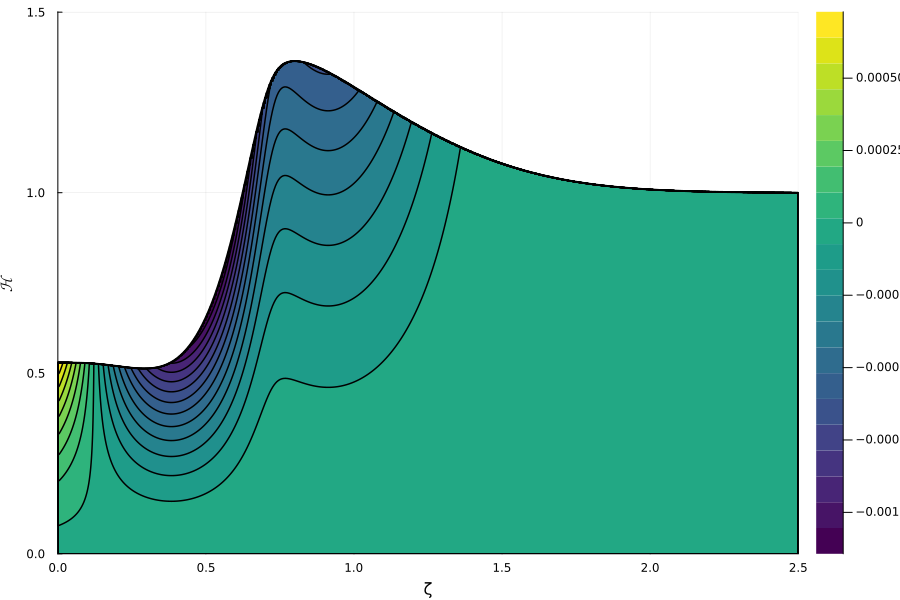}}
			\subfigure[$\mu = 0.1, \delta = 0.00$]{\label{fig:5_e}\includegraphics*[height=4.5cm, width=5.5cm]{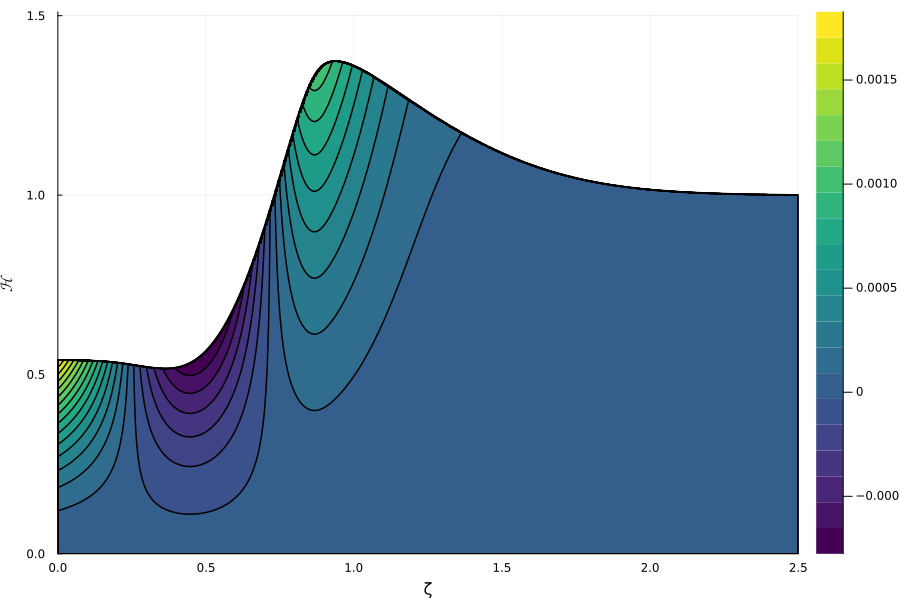}}
			\subfigure[$\mu = 0.05, \delta = 0.04$]{\label{fig:5_f}\includegraphics*[height=4.5cm, width=5.5cm]{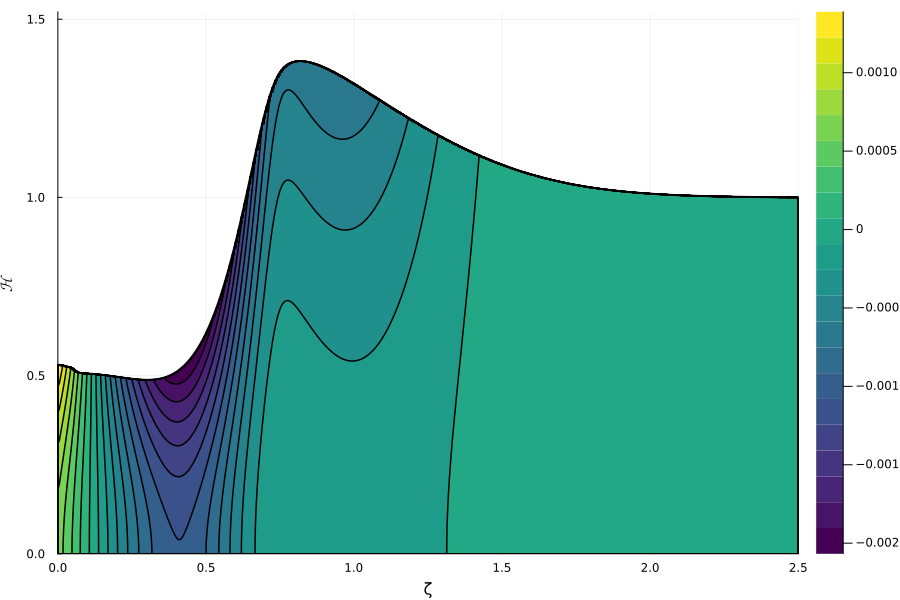}}
		\end{center} 
		\caption{ \scriptsize The streamfunction of the $\mathcal{H}$- profile for (a) $\mu = 0.0, \delta = 0.0$, (b) $\mu = 0.0, \delta = 0.04$, (c) $\mu = 0.0, \delta = 0.08$, (d) $\mu = 0.05, \delta = 0.0$, (e) $\mu = 0.1, \delta = 0.0$, (f) $\mu = 0.05, \delta = 0.04$ at time $\tau = 5.0$. The relevant parameters are $\beta_a = 1.0$, $Pe_s = 10.0$, $Pe_b = 10.0$, $\alpha = 1/7$, $\xi_m = 0.25$, $\xi_w = 0.1$. } \label{fig:5}
	\end{figure}
	
	The individual effect of the slip parameter $\delta$ in the absence of external shear $\mu$ on the streamfunction is portrayed in Figures \ref{fig:5_a} - \ref{fig:5_c}. It can be seen in all the figures that the closed recirculation vortex is created near the deposition of surfactant and bounded next to the region behind the capillary ridge. The viscous drag at the bottom wall in the case of no-slip forces the streamlines to attain a zero velocity and squeezes the vortex vertically. By contrast, the slippery substrate stimulates the streamlines near the bottom wall to gain momentum and the lower part of the vortex to slide along the slippery wall. The slippery walls help more fluid to slide along and help the shock fronts to gain volume and become elongated and shift forward. Moreover, the solitary effect of external shear $\mu$ in the absence of slip length $\delta$ on the streamfunction is shown in figures \ref{fig:5_d} - \ref{fig:5_e}. The streamlines clearly suggest that the external shear force inserts a uniform forward momentum along the streamwise direction across the entire interface. Also, the increasing external shear force pushes the recirculation vortex towards the front edge of the monomer, compacting it vertically. The shear force propels the top layer of the fluid so firmly that the vortex is pinned near the bottom layer and the streamlines near the recirculation zone become almost vertical (see Figure \ref{fig:5_e}). Next, the combined effect of external shear force and a slippery bottom on the streamfunction is explored in Figure \ref{fig:5_f}. In this scenario, the vortex at the recirculation zone is stretched out along the swelling as the acceleration due to the external shear along the interface and slide along the slippery bottom moves more fluid volume into the accelerating wedge as the viscous dissipation within the vortex reduces.

	\subsubsection{Front}
	If the supply rate of surfactant to a monolayer is chosen at $t^{1/2}$ instead of $t^{1/3}$, then the growth rate of monolayer length and shock becomes the same, and it continues to be so \cite{borgas1988monolayer, jensen1992insoluble, jensen1993spreading}. The transformation for this scenario can be considered as
	\bea
	\zeta = \displaystyle \frac{x}{t^{1/2}};~~\tau = t;~~\mathcal{H}(\zeta, \tau) = h(x,t);~~ \mathcal{F}(\zeta, \tau) =  \Gamma_0(x,t) \label{eqn29}
	\eea
	and the total mass of surfactant will be
	\bea
	\mathscr{M} = \mathcal{M} t^{1/2},~~\text{where}~~ \mathcal{M} = \displaystyle \int_{0}^{\infty} \left( \frac{\mathcal{H}}{\beta_a} + 1 \right) \mathcal{F} d\zeta.
	\eea
	With the transformations \eqref{eqn29}, the evolution equations \eqref{eqn16} \& \eqref{eqn22} reduces to
	
	\bea
	\displaystyle \tau \mathcal{H}_{\tau} - \frac{1}{2} \zeta \mathcal{H}_{\zeta} + \left[ \left(\mathcal{H}^2 + 2 \delta \mathcal{H} \right) \left\{ - \frac{3 (\alpha+1) \theta(\alpha) }{ (1 + \theta(\alpha) \mathcal{F})^4} \mathcal{F}_{\zeta} + \frac{\mu}{2} \tau^{1/2} \right\} \right]_{\zeta} = 0,\label{eqn31} \\
	\nonumber \displaystyle \mathcal{H} \left\{ \tau \mathcal{F}_{\tau} - \frac{1}{2}\zeta \mathcal{F}_{\zeta} + \left[ \left( - \frac{3 (\alpha + 1) \theta(\alpha) }{2 \left(1 + \theta(\alpha) \mathcal{F}\right)^4}  \mathcal{F}_{\zeta} + \frac{\mu}{2} \tau^{1/2} \right) \left( \mathcal{H} + 2 \delta \right) \right] \mathcal{F}_{\zeta} + \frac{\left( \mathcal{H} \mathcal{F}_{\zeta} \right)_{\zeta}}{\mathcal{H} Pe_b} \right\} \\
	+ \beta_a \left\{  \tau \mathcal{F}_{\tau} - \frac{1}{2}\zeta \mathcal{F}_{\zeta} + \left[ \left( - \frac{3 (\alpha + 1) \theta(\alpha) }{ \left(1 + \theta(\alpha) \mathcal{F}\right)^4} \mathcal{F}_{\zeta} + \mu \tau^{1/2} \right) \left(\mathcal{H} + \delta \right) \mathcal{F} \right]_{\zeta} + \frac{1}{Pe_s} \mathcal{F}_{\zeta \zeta} \right\} = 0. \label{eqn32}
	\eea
	Here, the diffusion terms have no explicit time dependence, so the equations \eqref{eqn31}-\eqref{eqn32} conserve the total mass of the surfactant $\mathcal{M}$ during the flow, with the availability of the upstream boundary condition
	\bea
	\displaystyle \mathcal{M} + 2 \mathcal{F}_{\zeta} \left[ \frac{\mathcal{H}}{\beta_a} \left( \frac{1}{Pe_b} + \frac{\mathcal{H F}}{2} \right) + \frac{1}{Pe_s} + \mathcal{H F} \right] = 0,~~\text{at}~~\zeta = 0. \label{eqn33}
	\eea
	
	The equations \eqref{eqn31}-\eqref{eqn32} were solved using the methods previously mentioned with the following initial conditions
	\bea
	\mathcal{H}(\zeta, 1) = 1~~\text{and}\\
	\displaystyle \zeta - \zeta_m + \frac{\zeta_m}{\zeta_a} \mathcal{F}\bigg|_{\tau = 1} + \nu \log( \mathcal{F}\bigg|_{\tau = 1}) = 0,~~\text{with}~~\nu \ll 1.
	\eea

	\begin{figure}[ht!]
		\begin{center} 
			\subfigure[$\delta = 0.00$]{\label{fig:6_a}\includegraphics*[height=4.5cm, width=5.5cm]{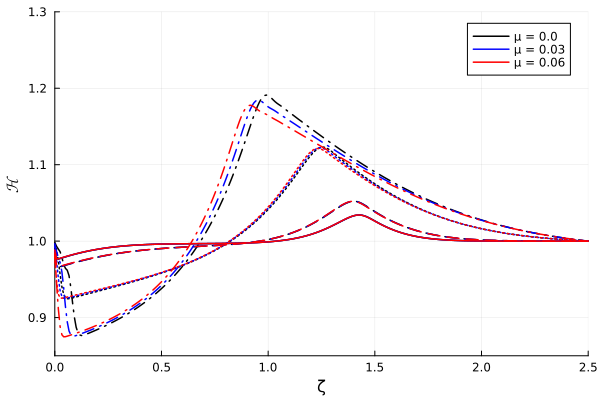}}
			\subfigure[$\delta = 0.04$]{\label{fig:6_b}\includegraphics*[height=4.5cm, width=5.5cm]{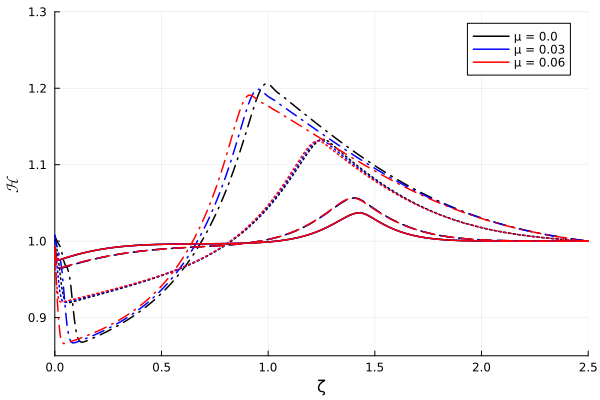}}
			\subfigure[$\delta = 0.08$]{\label{fig:6_c}\includegraphics*[height=4.5cm, width=5.5cm]{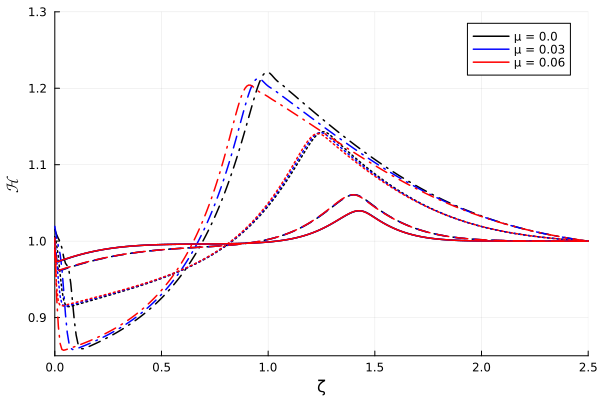}}
			\subfigure[$\delta = 0.00$]{\label{fig:6_d}\includegraphics*[height=4.5cm, width=5.5cm]{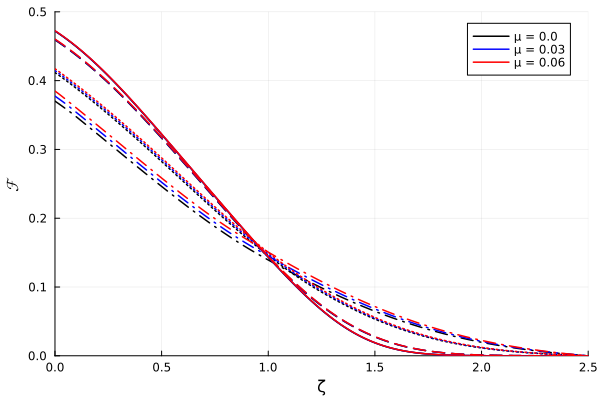}}
			\subfigure[$\delta = 0.04$]{\label{fig:6_e}\includegraphics*[height=4.5cm, width=5.5cm]{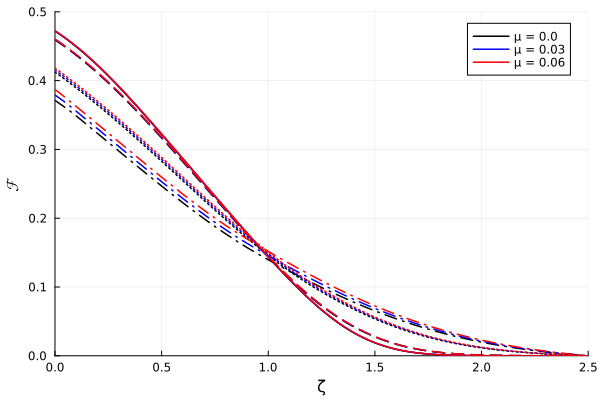}}
			\subfigure[$\delta = 0.08$]{\label{fig:6_f}\includegraphics*[height=4.5cm, width=5.5cm]{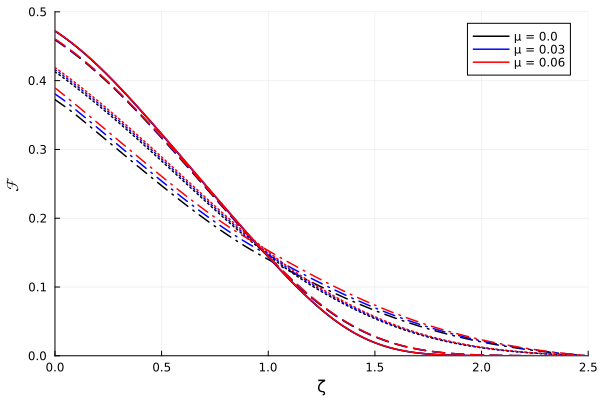}}
		\end{center} 
		\caption{ \scriptsize The variation (a) $\mathcal{H}$, and (b) $\mathcal{F}$ profile for a soluble surfactant-laden spreading front in the case of fast sorption kinetics by varying the externally imposed shear force. The solid, dashed, dotted, and dash-dot curves correspond to $\tau = 1.05,~ 1.1,~ 1.5,~ 3.0$. The relevant parameters are $\beta_a = 1.0$, $Pe_s = 10.0$, $Pe_b = 1.0$, $\alpha = 1/7$, $\xi_m = 1.36$, $\xi_a = 0.5$, $\nu = 0.02$. } \label{fig:6}
	\end{figure}
	
	Figures (\ref{fig:6}) illustrate the combined effect of external shear force $\mu$ and slip parameter $\delta$ on the film thickness $\mathcal{H}$ and surfactant concentration $\mathcal{F}$ in the case of front-type surfactant distribution. In this kind of distribution, a localized Marangoni stress is created at the leading edge, which drives fluid from the contaminated to the clean region, generating a hump, which is followed by a depressed thinning region due to the drainage of fluid to the hump (see figure \ref{fig:6_a}). In the absence of a slippery bottom (i.e., $\delta = 0.0$), the increasing external force $\mu$ pushes the hump front downstream, and the steep slope of the upstream gets even more elevated. But, as slip length is introduced and increased, the hump gets even taller and advances much more rapidly, as can be seen from figures \ref{fig:6_a}-\ref{fig:6_c}. Physically, a slippery bottom relaxes the wall friction, which helps fluid to accelerate faster, and conservation of surfactant mass ensures the positive gradient along the free surface, which generates the hump. This hump gains height due to the more fluid sliding along the slippery substrate and the presence of external shear that pushes the advancing shock front. Furthermore, the convective velocity of the surfactant monolayer mainly depends on the surface velocity. As slippery bottom $\delta$ and external shear $\mu$ accelerate the surface advection \cite{paul2025stability}, the surfactant front is transported at an amplified velocity and turns into a linear profile over a large time (see figures \ref{fig:6_d} - \ref{fig:6_f}).

	\begin{figure}[ht!]
		\begin{center} 
			\subfigure[$\mu = 0.00$]{\label{fig:7_a}\includegraphics*[height=4.5cm, width=5.5cm]{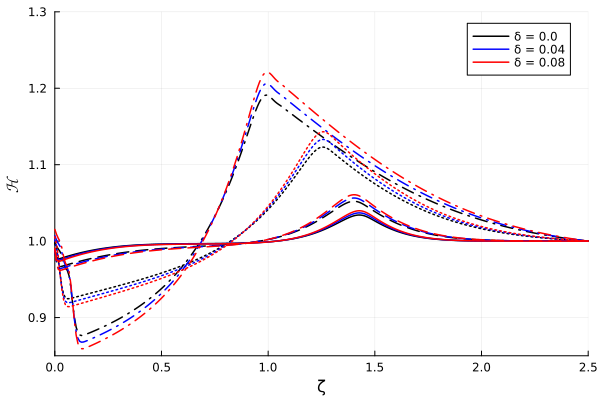}}
			\subfigure[$\mu = 0.03$]{\label{fig:7_b}\includegraphics*[height=4.5cm, width=5.5cm]{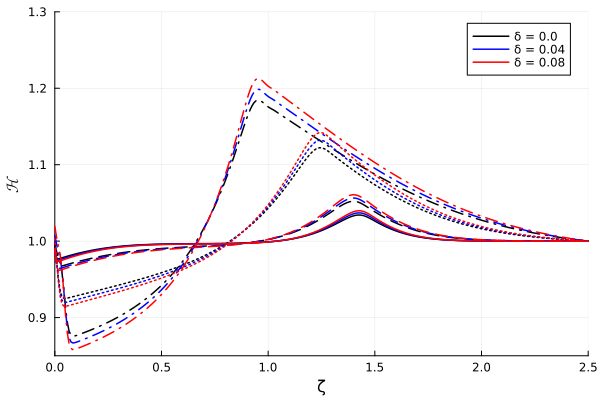}}
			\subfigure[$\mu = 0.06$]{\label{fig:7_c}\includegraphics*[height=4.5cm, width=5.5cm]{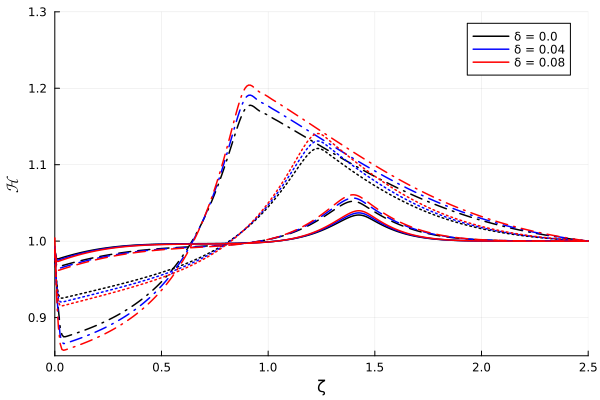}}
			\subfigure[$\mu = 0.00$]{\label{fig:7_d}\includegraphics*[height=4.5cm, width=5.5cm]{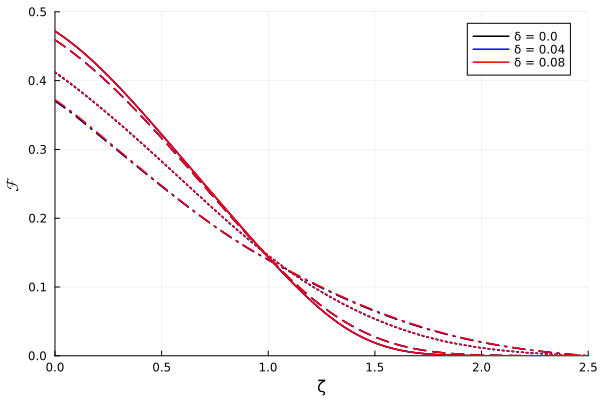}}
			\subfigure[$\mu = 0.03$]{\label{fig:7_e}\includegraphics*[height=4.5cm, width=5.5cm]{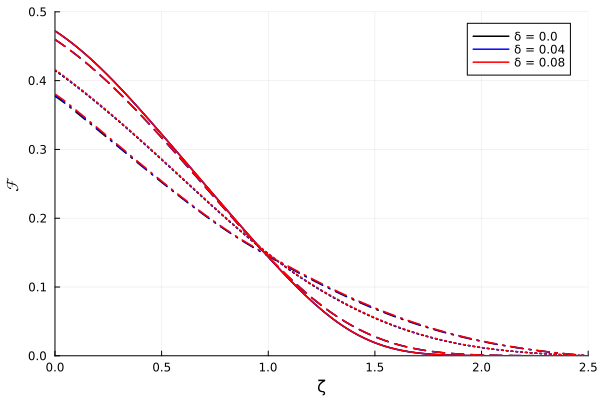}}
			\subfigure[$\mu = 0.06$]{\label{fig:7_f}\includegraphics*[height=4.5cm, width=5.5cm]{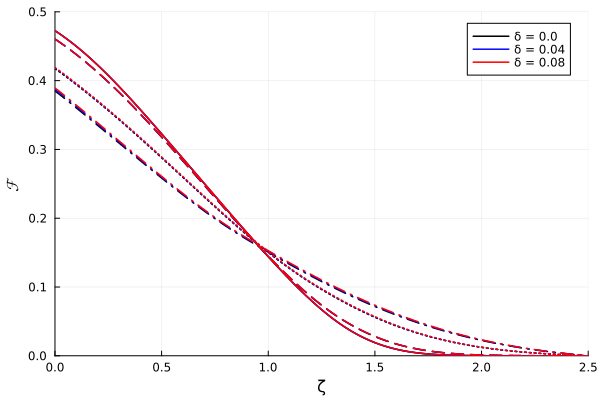}}
		\end{center} 
		\caption{ \scriptsize The variation (a) $\mathcal{H}$, and (b) $\mathcal{F}$ profile for a soluble surfactant-laden spreading front in the case of fast sorption kinetics by varying the slip parameter. The solid, dashed, dotted, and dash-dot curves correspond to $\tau = 1.05,~ 1.1,~ 1.5,~ 3.0$. The relevant parameters are $\beta_a = 1.0$, $Pe_s = 10.0$, $Pe_b = 1.0$, $\alpha = 1/7$, $\xi_m = 1.36$, $\xi_a = 0.5$, $\nu = 0.02$. } \label{fig:7}
	\end{figure}
	
	The figures (\ref{fig:7}) are plotted for different external shear forces $\mu$ by varying slip length $\delta$ in the case of front-type surfactant deposition. In each panel \ref{fig:7_a} - \ref{fig:7_c}, it can be observed that as slip length $\delta$ is increased, the hump in the film thickness gets systematically taller. When $\delta = 0.0$, the viscous drag resists the fluid flow; a positive slip length ($\delta > 0$) allows more volume of fluid to flux through, and this extra volume of fluid piles up in the pulse and increases its height. However, the increased external shear mobilizes the interface fluid and reduces the time for the extra volume of fluid (due to the slip bottom) to pile up in the pulse. The surfactant concentration $\mathcal{F}$  is impacted by external shear as it pushes the interface; the monolayer is also swept, which tends to increase the surfactant concentration for higher external shear force $\mu$ (see figures \ref{fig:7_d} - \ref{fig:7_f}).

	\subsection{Slow sorption kinetics $(k_a \ll 1)$}
	
	\subsubsection{Strip}
	
	This subsection is dedicated to deciphering the transient phase due to the slow sorption kinetics for the case of the spreading strip. So, it is convenient to transform the equations \eqref{eqn18}-\eqref{eqn20} with the following 
	\bea
	\zeta = \displaystyle \frac{x}{t^{1/3}};~~\tau = t;~~\mathcal{H}(\zeta, \tau) = h(x,t);~~ \mathcal{F}(\zeta, \tau) = t^{1/3} \Gamma_0(x,t);~~\mathcal{G}(\zeta, \tau) = t^{1/3} \Gamma_a(x,t);
	\eea
	rules, which result in
	\bea
	\displaystyle \tau \mathcal{H}_{\tau} - \frac{1}{3} \zeta \mathcal{H}_{\zeta} + \left[ \left( \mathcal{H}^2 + 2 \delta \mathcal{H} \right) \left( - \frac{3 (\alpha + 1) \theta(\alpha) \tau^{1/3}}{2 (\tau^{1/3} + \theta(\alpha) \mathcal{G})} \mathcal{G}_{\zeta} + \frac{\mu}{2} \tau^{2/3} \right) \right]_{\zeta} = 0,\label{eqn37} \\
	\nonumber \displaystyle \tau \mathcal{G}_{\tau} - \frac{1}{3}(\zeta \mathcal{G})_{\zeta} + \left[ \left( - \frac{3 (\alpha + 1) \theta(\alpha) \tau^{4/3}}{ (\tau^{1/3} + \theta(\alpha) \mathcal{G})} \mathcal{G}_{\zeta} + \mu \tau^{2/3} \right) \left( \mathcal{H} + \delta \right) \mathcal{G} \right]_{\zeta} - \frac{1}{Pe_s} \tau^{1/3} \mathcal{G}_{\zeta \zeta}\\
	- k_a \left[ \mathcal{F} \left( \tau - \tau^{2/3} \mathcal{G} \right) - \tau \mathcal{G} \right] = 0,\label{eqn38}\\
	\nonumber \displaystyle \tau \mathcal{F}_{\tau} - \frac{1}{3}(\zeta \mathcal{F})_{\zeta} + \left[ - \frac{3 (\alpha + 1) \theta(\alpha) \tau^{4/3}}{ 2 (\tau^{1/3} + \theta(\alpha) \mathcal{G})} \mathcal{G}_{\zeta} + \frac{\mu}{2} \tau^{2/3} \right] \left( \mathcal{H} + 2 \delta \right) \mathcal{F}_{\zeta} - \frac{1}{\mathcal{H} Pe_b} \tau^{1/3} \left( \mathcal{H} \mathcal{F}_{\zeta} \right)_{\zeta}\\
	- \frac{1}{\mathcal{H}} \beta_a k_a \left[ \mathcal{F} \left( \tau - \tau^{2/3} \mathcal{G} \right) - \tau \mathcal{G} \right] = 0.\label{eqn39}
	\eea
	These equations are solved with the initial conditions
	\bea
	\nonumber \mathcal{H}(\zeta, 1) = 1.0;\\
	\mathcal{F}(\zeta, 1) = 0.0;\label{eqn40}\\
	\nonumber \mathcal{G}(\zeta, 1) = \displaystyle \frac{1}{2} \left[ 1 - \tanh(\frac{\zeta - 0.5}{0.1}) \right].
	\eea

	\begin{figure}[ht!]
		\begin{center} 
			\subfigure[$\delta = 0.00$]{\label{fig:8_a}\includegraphics*[height=4.5cm, width=5.5cm]{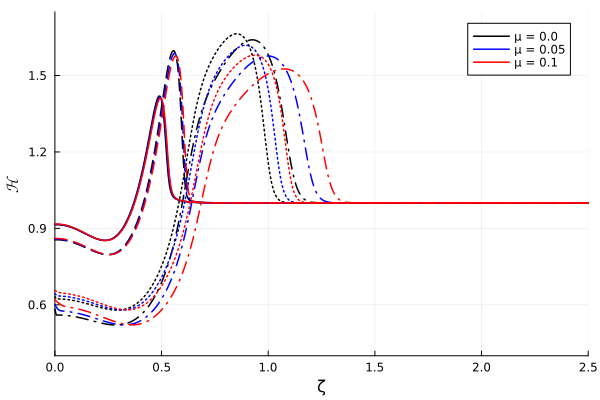}}
			\subfigure[$\delta = 0.04$]{\label{fig:8_b}\includegraphics*[height=4.5cm, width=5.5cm]{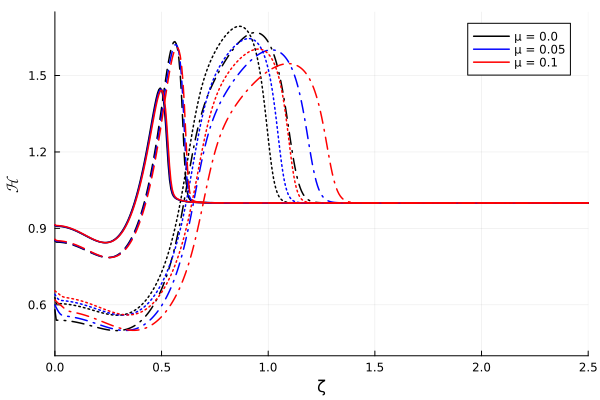}}
			\subfigure[$\delta = 0.08$]{\label{fig:8_c}\includegraphics*[height=4.5cm, width=5.5cm]{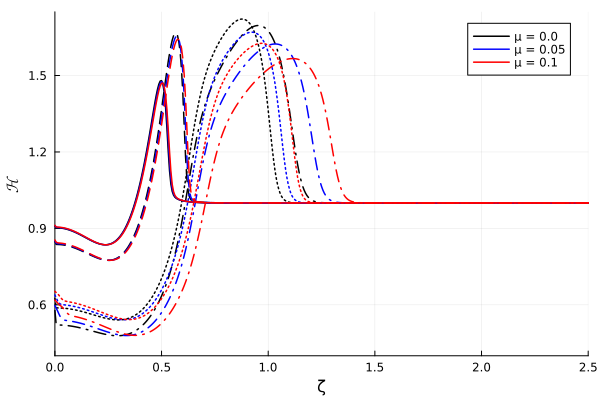}}
			\subfigure[$\delta = 0.00$]{\label{fig:8_d}\includegraphics*[height=4.5cm, width=5.5cm]{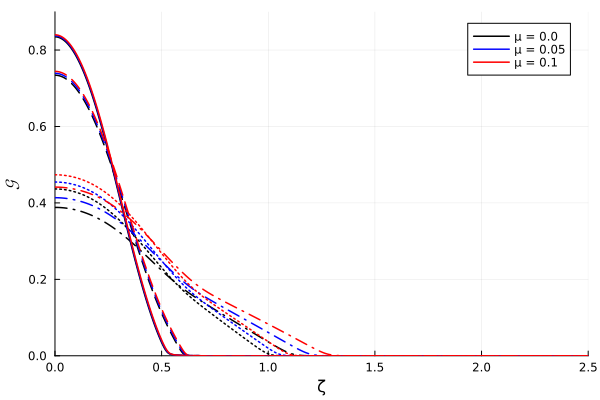}}
			\subfigure[$\delta = 0.04$]{\label{fig:8_e}\includegraphics*[height=4.5cm, width=5.5cm]{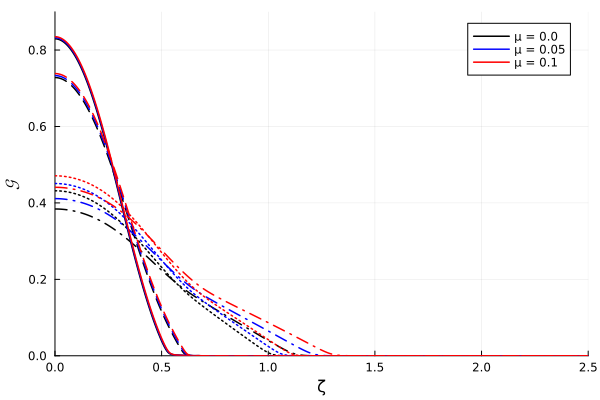}}
			\subfigure[$\delta = 0.08$]{\label{fig:8_f}\includegraphics*[height=4.5cm, width=5.5cm]{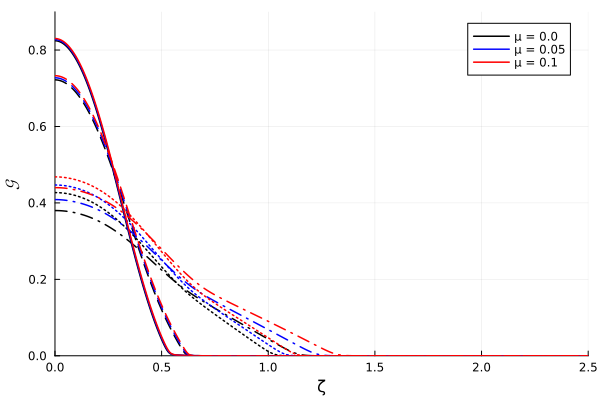}}
			\subfigure[$\delta = 0.00$]{\label{fig:8_g}\includegraphics*[height=4.5cm, width=5.5cm]{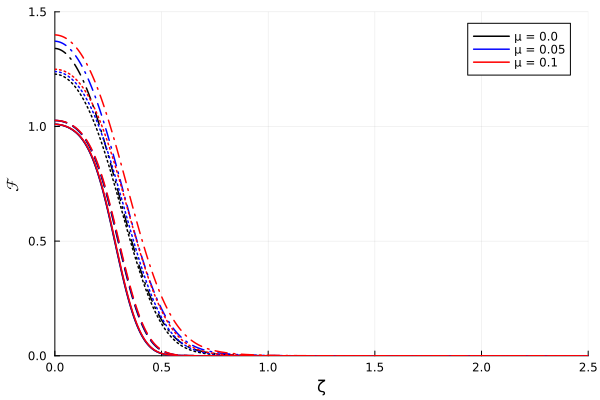}}
			\subfigure[$\delta = 0.04$]{\label{fig:8_h}\includegraphics*[height=4.5cm, width=5.5cm]{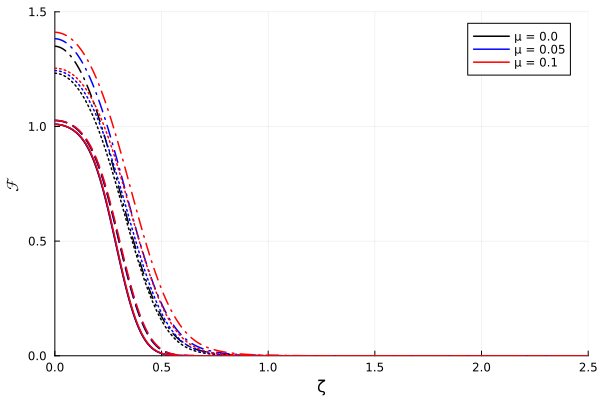}}
			\subfigure[$\delta = 0.08$]{\label{fig:8_i}\includegraphics*[height=4.5cm, width=5.5cm]{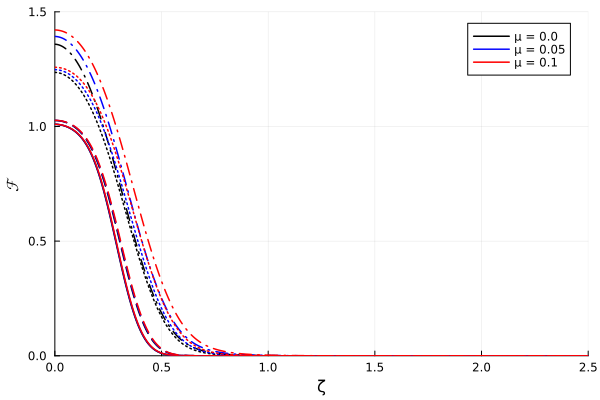}}
		\end{center} 
		\caption{ \scriptsize The variation (a)-(c) $\mathcal{H}$, (d)-(f) $\mathcal{G}$, and (g)-(i) $\mathcal{F}$ profile for a soluble surfactant-laden spreading strip in case of slow sorption kinetics by varying the externally imposed shear force. The solid, dashed, dotted, and dash-dot curves correspond to $\tau = 1.05,~ 1.1,~ 2.0,~ 3.0$. The relevant parameters are $\beta_a = 1.0$, $Pe_s = 100.0$, $Pe_b = 100.0$, $\alpha = 1/7$, $\xi_m = 0.25$, $\xi_w = 0.1$, $k_a = 1.0$, $R_a = 1.0$. } \label{fig:8}
	\end{figure}
	
	The equations \eqref{eqn37}-\eqref{eqn39} are solved using the initial conditions at \eqref{eqn40} and illustrated in the figures (\ref{fig:8}). It is important to note that in this case, the flux term is present and will play an important role in spreading dynamics. Here, the timescale of desorption is taken to be slow, so at the beginning, the distributed surfactant behaves more as an insoluble one. This can be observed from Figure \ref{fig:8_a}, where at the initial stages, i.e., $\tau = 1.05$ and $\tau = 1.1$, the film thickness displays a sharp shock at the surfactant deposition. As time passes, the desorption takes place in a transient period, and at this time, the surfactant dissolves into the bulk from the interface, which is observable from figures \ref{fig:8_d} and \ref{fig:8_g}, as the concentration of $\mathcal{G}$ decreases and $\mathcal{F}$ increases at the later part of the simulated results (see figures \ref{fig:8_d} and \ref{fig:8_g} at time $\tau = 2.0$ and $\tau = 3.0$). Moreover, as desorption happens, the amount of surfactant at the interface reduces, which relaxes the local Marangoni stress that consequently slows down the spreading rate. Now, the effect of the external shear force $\tau$ for different slip lengths $\delta$ is portrayed in the figures (\ref{fig:8}). These figures clearly show that as the external shear force is in effect, it boosts the driving force of the fluid behind the leading edge, which leads to an additional thinning of the fluid thickness. This reduction in fluid volume amplifies the local surfactant concentration $\mathcal{F}$, which promotes the adsorption of surfactant from the bulk to the interface and eventually reaches equilibrium. Also, as time progresses, the interfacial surfactant concentration reduces due to desorption, which moderates the local Marangoni stress and converts the film thickness from shock to pulse (see figure \ref{fig:8_a}). Furthermore, the external shear force $\mu$ reduces the pulse height systematically. Meanwhile, the slippery bottom helps the bulk fluid to move faster and also the bulk and interface surfactant to maintain the adsorption-desorption rates by reducing the concentration lag (see figures \ref{fig:8_a} - \ref{fig:8_i}).

	\begin{figure}[ht!]
		\begin{center} 
			\subfigure[$\mu = 0.00$]{\label{fig:9_a}\includegraphics*[height=4.5cm, width=5.5cm]{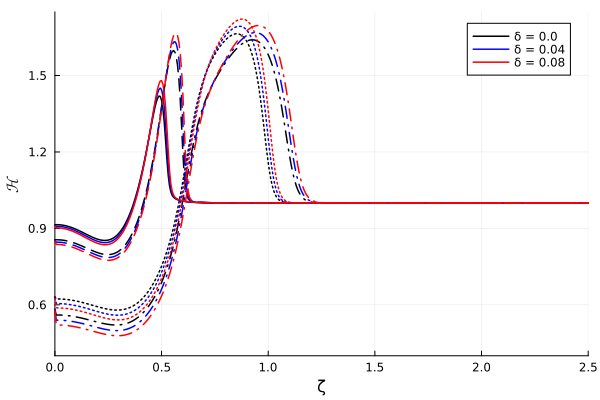}}
			\subfigure[$\mu = 0.05$]{\label{fig:9_b}\includegraphics*[height=4.5cm, width=5.5cm]{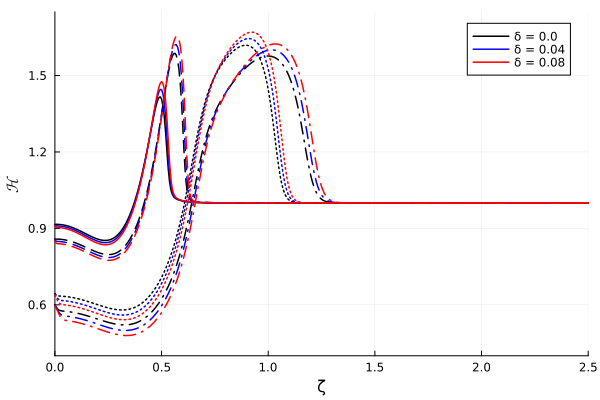}}
			\subfigure[$\mu = 0.10$]{\label{fig:9_c}\includegraphics*[height=4.5cm, width=5.5cm]{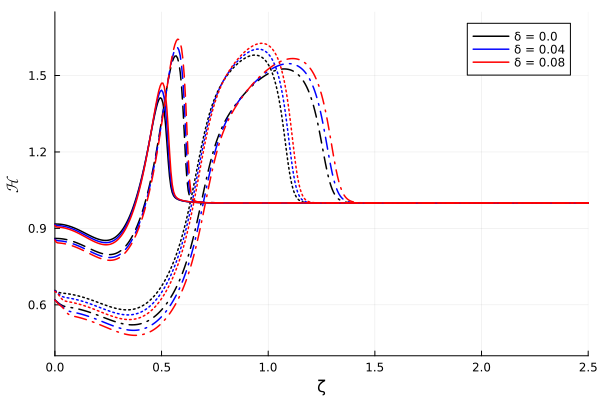}}
			\subfigure[$\mu = 0.00$]{\label{fig:9_d}\includegraphics*[height=4.5cm, width=5.5cm]{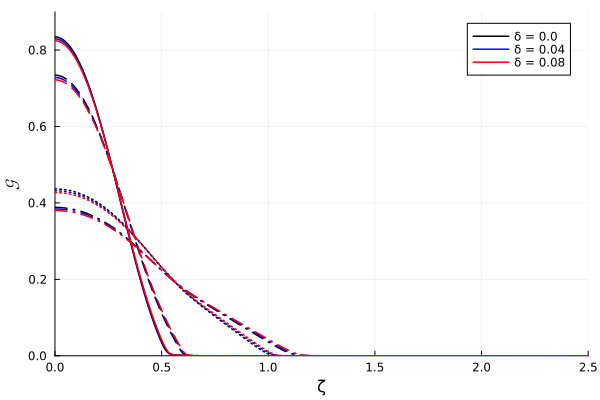}}
			\subfigure[$\mu = 0.05$]{\label{fig:9_e}\includegraphics*[height=4.5cm, width=5.5cm]{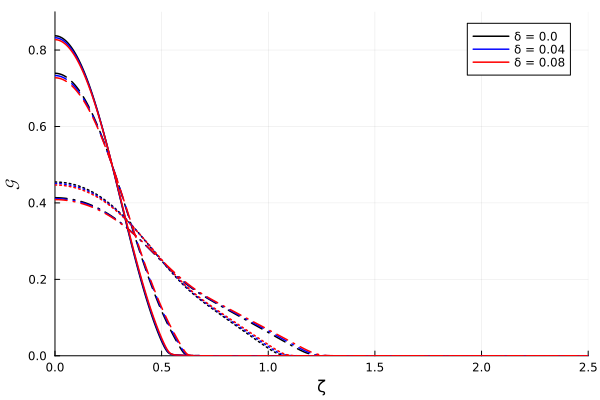}}
			\subfigure[$\mu = 0.10$]{\label{fig:9_f}\includegraphics*[height=4.5cm, width=5.5cm]{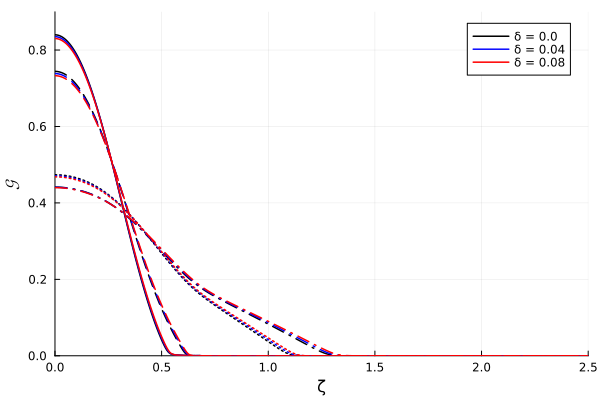}}
			\subfigure[$\mu = 0.00$]{\label{fig:9_g}\includegraphics*[height=4.5cm, width=5.5cm]{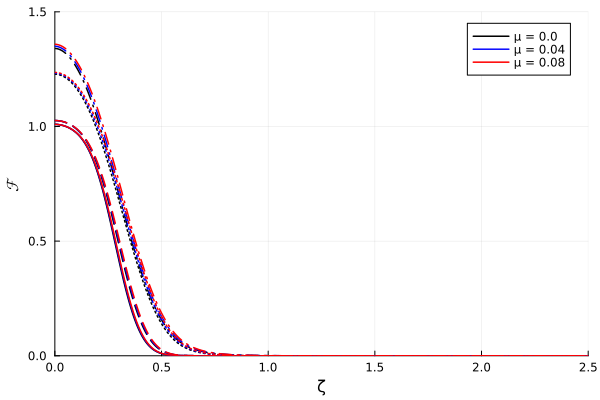}}
			\subfigure[$\mu = 0.05$]{\label{fig:9_h}\includegraphics*[height=4.5cm, width=5.5cm]{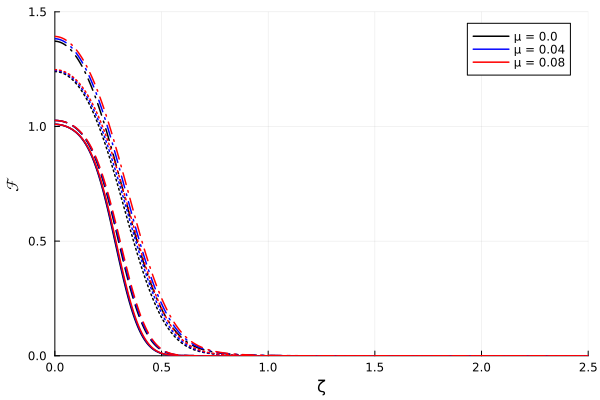}}
			\subfigure[$\mu = 0.10$]{\label{fig:9_i}\includegraphics*[height=4.5cm, width=5.5cm]{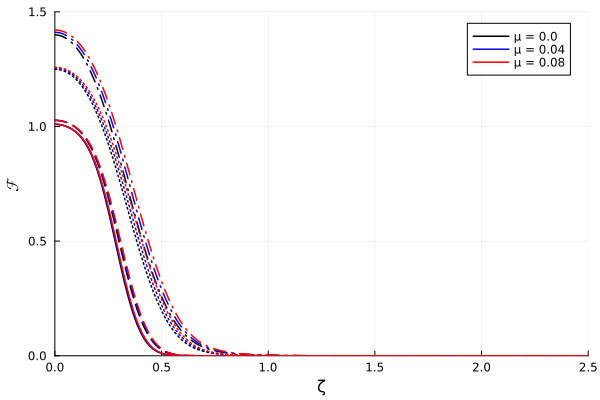}}
		\end{center} 
		\caption{ \scriptsize The variation (a) $\mathcal{H}$, (b) $\mathcal{G}$, and (c) $\mathcal{F}$ profile for a soluble surfactant-laden spreading strip in case of slow sorption kinetics by varying the slip parameter. The solid, dashed, dotted, and dash-dot curves are corresponding to $\tau = 1.05,~ 1.1,~ 2.0,~ 3.0$. The relevant parameters are $\beta_a = 1.0$, $Pe_s = 100.0$, $Pe_b = 100.0$, $\alpha = 1/7$, $\xi_m = 0.25$, $\xi_w = 0.1$, $k_a = 1.0$. } \label{fig:9}
	\end{figure}
	
	Figures (\ref{fig:9}) demonstrate the impact of the slip parameter on the film thickness $\mathcal{H}$, surfactant concentrations $\mathcal{F}$ and $\mathcal{G}$ for some fixed external shear force $\mu$. In case of no slip (see figure \ref{fig:9_a}),  the viscous drag restricts the fluid volume to be drawn by the Marangoni gradient, which creates the shock, just like the case of strip-type distribution of rapid sorption kinetics; and in a similar way, when the slip parameter is introduced, more fluid passes through and accumulates at the pulse and helps the pulse gain the extra bit of volume because of the slippery bottom. More so, due to the assumption of a cross-sectional average, squeezing of fluid film means the concentration of the bulk surfactant rises and it reaches up to a point where the concentration of bulk surfactant is more than that of the interface surfactant, which reverses the flux (see figures \ref{fig:9_d}-\ref{fig:9_i}). As the introduction of a slippery bottom allows more fluid flow, so does more surfactant concentration spread; hence, the concentration $\mathcal{G}$ of interface surfactant reduces. On the contrary, the external shear force $\mu$ pushes the fluid and stretches the film horizontally, which acts as a suppression to reduce the pulse height (see figure \ref{fig:9_a}-\ref{fig:9_c}).
	
	\section{Rupture dynamics}\label{sec:rup}
	The above discussion about surfactant-induced spreading dynamics clearly indicates thinning of the film underneath the surfactant monolayer, which can lead to film rupture. Although \citet{jensen1992insoluble} mentioned that Marangoni and viscous forces are not sufficient to instigate film rupture in finite time, several experimental and theoretical \cite{thete2015self, duran2019instability, moreno2020stokes, sprittles2023rogue} explorations suggest otherwise. It can be described as due to surfactant deposition, the thin film experiences thinning because of the Marangoni forces, and when the film thickness reaches below $1000$ \AA, van der Waals forces become effective. However, the van der Waals forces reduce the film thickness to zero at some point, which eventually leads to a rupture phenomenon.
	
	\subsection{Linear stability analysis}
	
	This section is dedicated to examining the linear stability analysis of the one-dimensional evolution equations, which describe the rupture of the soluble surfactant-laden thin film flow over an inclined plane in the presence of external shear. The aim is to observe the combined influence of external shear force and slip parameter on the rupture of the thin film. Infinitesimal perturbations to an initially ruptured thin film are considered in normal mode form in terms of complex frequency $\omega$ and wavenumbers $k$ as
	
	\bea
	& \nonumber h(x,t) = H_b + \widetilde{h}_1 \exp(\mathrm{i}k x + \omega t),\\
	& \Gamma_a(x,t) = \Gamma_{a,B} + \widetilde{\Gamma}_{a,1} \exp(\mathrm{i}k x + \omega t), \label{eqn41}\\
	& \nonumber \Gamma_0(x,t) = \Gamma_{0,B} + \widetilde{\Gamma}_{0,1} \exp(\mathrm{i}k x + \omega t). 
	\eea
	
	Now, the equations \eqref{eqn41} are substituted into the equations \eqref{eqn16}, \eqref{eqn20}, and \eqref{eqn21}, where $H_b$ denotes the uniform thickness, $\Gamma_{a,B}$, $\Gamma_{0,B}$ are interface and bulk surfactant concentrations at equilibrium, respectively. Moreover, the terms with tilde notations denote the amplitude of the infinitesimal perturbation. The linear terms of the perturbations are only considered here for the linear stability analysis. Hence, the following homogeneous system of equations can be obtained:
	
	\bea
	\mathcal{N X} = \mathbf{0} \label{eqn42}
	\eea
	where $\mathcal{N}$ is given by
	
	\begin{align}
		\hspace{-1.75cm}
		\left(
		\begin{array}{@{}c@{\hspace{3mm}}c@{\hspace{3mm}}c@{}}
			\begin{aligned}
				& \nonumber i Bo H_b k (H_b+\delta )\\
				& \nonumber +\frac{k^2 \left(-3 A+C H_b^4 k^2\right) (2 H_b+3 \delta )}{6 H_b^2}\\
				& \nonumber +i k (H_b+\delta) \mu +\omega
			\end{aligned}  & \begin{aligned}
				& \nonumber \frac{3 H_b k^2 (1+\alpha ) \left(1+\left(\frac{1+\alpha }{\alpha }\right)^{1/3}\right) (H_b+2 \delta )}{2 \left(1+\Gamma_{aB}+\left(\frac{1+\alpha }{\alpha }\right)^{1/3}  \Gamma_{a,B}\right)^4}
			\end{aligned} & 0 \\[1.5cm]
			\begin{aligned}
				& \nonumber i Bo H_b k  \Gamma_{a,B}\\
				& \nonumber +\frac{k^2 \left(-3 A+C H_b^4 k^2\right)  \Gamma_{a,B}}{2 H_b^2}\\
				& \nonumber +i k  \Gamma_{a,B}
				\mu
			\end{aligned}  & \begin{aligned}
				& \nonumber k_a+\frac{k^2}{Pe_{s}}+k_a    \Gamma_{0,B}\\
				& \nonumber +\frac{3 k^2 (1+\alpha ) \left(1+\left(\frac{1+\alpha }{\alpha }\right)^{1/3}\right)
					\Gamma_{a,B} (H_b+\delta )}{\left(1+ \Gamma_{a,B}+\left(\frac{1+\alpha }{\alpha }\right)^{1/3}  \Gamma_{a,B}\right)^4}\\
				& \nonumber +i k \left(\frac{Bo H_b^2}{2}+(H_b+\delta ) \mu \right)+\omega
			\end{aligned}  & k_a  -k_a    \Gamma_{a,B} \\[2.5cm]
			\begin{aligned}
				& \nonumber \frac{k_a \beta_a  \Gamma_{0,B}}{H_b^2} -\frac{k_a  \beta_a  \Gamma_{a,B}}{H_b^2}\\
				& \nonumber -\frac{k_a    \beta_a  \Gamma_{0,B}  \Gamma_{a,B}}{H_b^2}
			\end{aligned} & \displaystyle \frac{k_a  \beta_a}{H_b}+\frac{k_a   		 \beta_a  \Gamma_{0,B}}{H_b} & \begin{aligned}
				&  \nonumber \displaystyle -\frac{k^2}{Pe_{b}}-\frac{k_a    \beta_a}{H_b}+\frac{k_a    \beta_a  \Gamma_{a,B}}{H_b}\\
				& \nonumber \hspace{-0.25cm} +\frac{1}{6} \left(-2 i Bo H_b^2 k-3 i Bo H_b k \delta -6 i H_b k \mu -6 i k \delta  \mu -6 \omega \right)
			\end{aligned} \\
		\end{array}
		\right)
	\end{align}
	
	and $\mathcal{X} = [\widetilde{h}_1~ \widetilde{\Gamma}_{a,1}~ \widetilde{\Gamma}_{0,1}]^{T}$.
	
	Now, the non-trivial solutions of the system of equations \eqref{eqn42} exist only if $\det(\mathcal{N}) = 0$. Using this condition, the dispersion relation is obtained as
	
	\bea
	\omega ^3 - a_1 \omega ^2 - a_2 \omega + a_3 = 0\label{eqn43}
	\eea
	
\begin{figure}[ht!]
		\begin{center} 
			\subfigure[]{\label{fig:10_a}\includegraphics*[height=6.0cm, width=8.5cm]{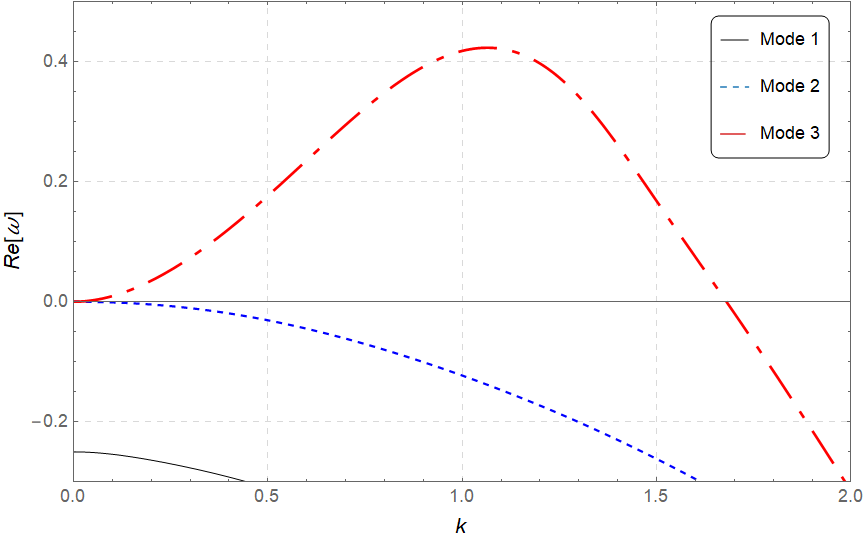}}
			\subfigure[]{\label{fig:10_b}\includegraphics*[height=6.1cm, width=8.5cm]{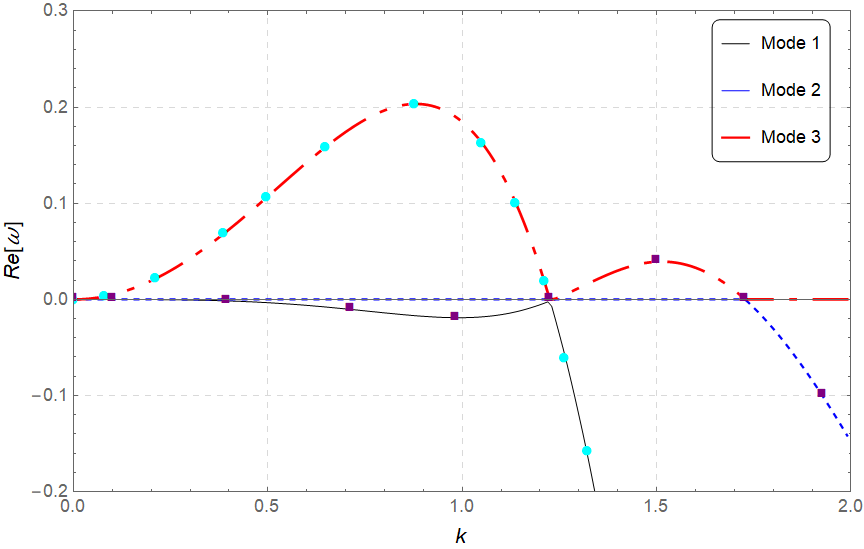}}
		\end{center}
		\caption{ \scriptsize (a) The real part of perturbation growth rate $Re[\omega]$ against wavenumbers $k$ by solving the dispersion relation \eqref{eqn43}. The parameters chosen are $Bo = 1.0,~H_b = 1.0,~k_a = 0.1,~Pe_b = 10.0,~Pe_{s} = 100, ~\Gamma_{0B} = 1.0,~\beta_a = 1.0,~\Gamma_{aB} = 0.5,~\mathcal{A} = 1.0,~\mathcal{C} = 1.0,~\alpha = 100, ~\mu=\delta=0.0$. (b) The perturbation growth rate $Re[\omega]$ against wavenumbers $k$ in the insoluble surfactant limit. $Bo = 2.0,~H_b = 1.0,~k_a = 0.0,~Peb\rightarrow \infty,~Pe_{s} = 1000, ~\Gamma_{0B} = 1.0,~\beta_a = 1.0,~\Gamma_{aB} = 0.5,~\mathcal{A} = 1.0,~\mathcal{C} = 1.0,~\mu=\delta=0.0$. The cyan circular and violet square points are the data points extracted from \citet{vivek2024rupture}. The cyan circular points refer to mode one, and the violet square points denote another mode in Figure 3(a) of their paper\cite{vivek2024rupture}.}  \label{fig:10}
	\end{figure}

	This equation is obtained and solved using ``MATHEMATICA'' software, and the solution is given in Appendix \ref{sec:apndx}. The complex frequency can be written as $\omega = Re[\omega] + \mathrm{i} Im[\omega]$, where $Re[\omega]$ represents the growth rate of perturbation, and $Im[\omega]/k$ is the phase velocity. If the real part of the perturbation growth rate $Re[\omega] > 0$, then the film is unstable, and if $Re[\omega] < 0$, then the film is stable. The equation \eqref{eqn43} has three roots or modes, and the real part of those ($Re[\omega]$) is plotted against wavenumbers $k$ in Figure \ref{fig:10_a} for fixed values of all the remaining parameters. Figure \ref{fig:10_a} illustrates that only one root has a positive real part and the other two have a negative real part. So, only one mode is unstable in this parameter range. The rest of the article will focus solely on this mode. However, in the insoluble surfactant limit, the behaviour of the other two modes becomes interesting, which is explored in the next Figure \ref{fig:10_b}. 
	
	The modes are validated in the insoluble surfactant limit with the results from Figure 3(a) of \citet{vivek2024rupture} (see Figure \ref{fig:10_b}). In the insoluble surfactant limit, the modes obtained from the dispersion relation \eqref{eqn43} are coalesced. Also, the data points of  \textit{mode 1} from \citet{vivek2024rupture} lie on two modes obtained by the current article, as can be observed in Figure \ref{fig:10_b}. On the other hand, \textit{mode 2} of \citet{vivek2024rupture} lies on three modes obtained by the current article. Next, the effect of external shear force $\mu$ and slip length $\delta$ on the linear stability is explored in the subsequent subsections.
	
	\subsubsection{Influence of external shear force $\mu$}
	
	\begin{figure}[ht!]
		\begin{center} 
			\subfigure[$Bo = 0.0$]{\label{fig:11_a}\includegraphics*[height=4.5cm, width=5.5cm]{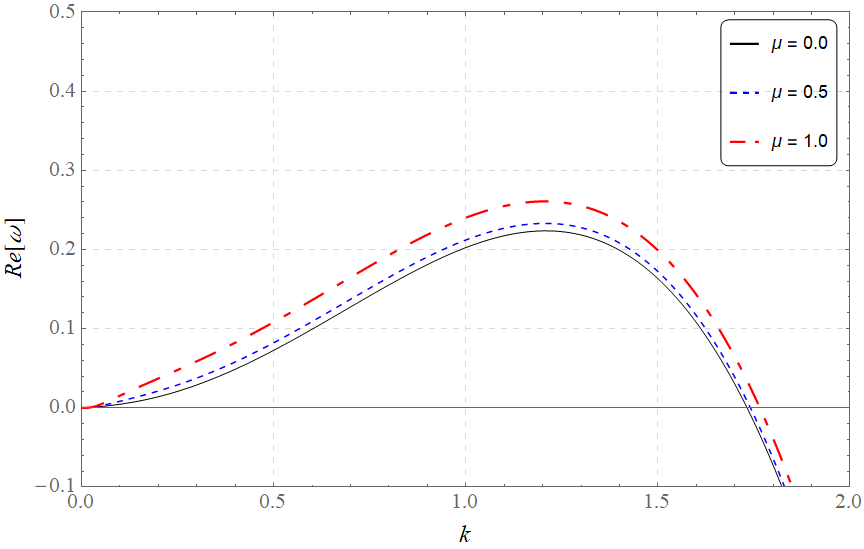}}
			\subfigure[$Bo = 1.0$]{\label{fig:11_b}\includegraphics*[height=4.5cm, width=5.5cm]{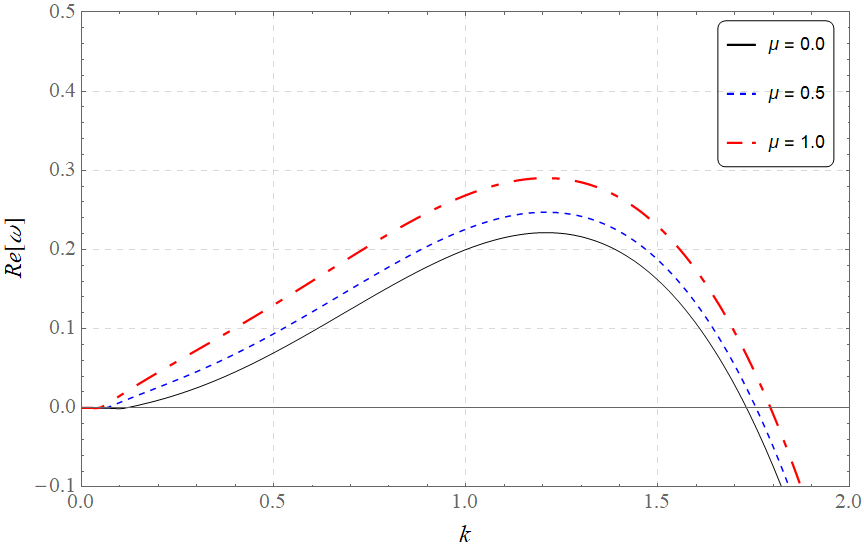}}
			\subfigure[$Bo = 10.0$]{\label{fig:11_c}\includegraphics*[height=4.5cm, width=5.5cm]{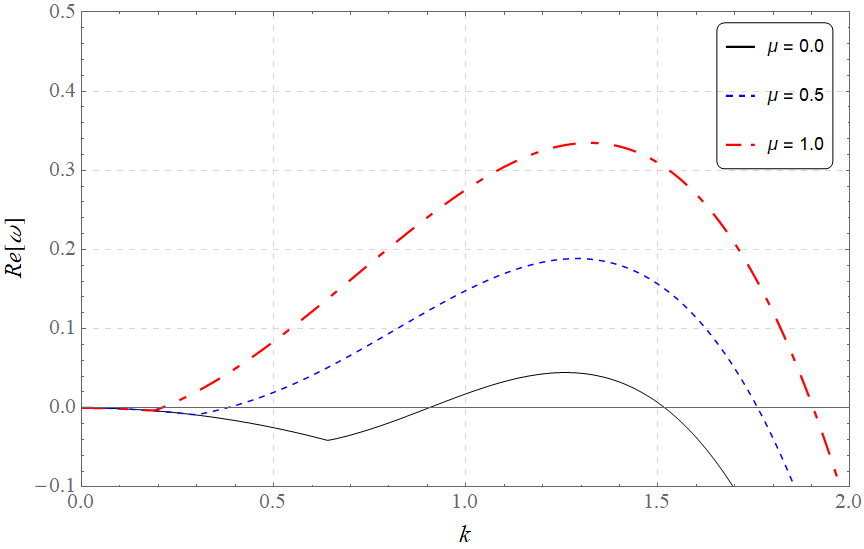}}
		\end{center} 
		\caption{ \scriptsize The growth rate curves corresponding to the dispersion relation against wavenumbers $k$ for different external shear force $\mu$ in case of (a) $Bo = 0.0$, (b) $Bo = 1.0$, and (c) $Bo = 10.0$. The rest of the parameters are $H_b = 1.0$, $\Gamma_{0,B} = 1.0$, $\Gamma_{a,B} = 1.0$, $k_a = 0.01$, $Pe_b = 10$, $Pe_{s} = 100$, $\beta_a = 1.0$, $\mathcal{A} = 1.0$, $\mathcal{C} = 1.0$, $\alpha = 0.01$, $\delta = 0.05$. } \label{fig:11}
	\end{figure}
	
	Figure \ref{fig:11_a} depicts that, when the Bond number is zero, the system becomes more unstable if the external shear force increases. Although in the case of a positive Bond number, the system tends to remain stable in the long-wave zone and the amplitude of the Bond number influences the range for a stable wavenumber regime, external shear force $\mu$ destabilizes the flow, as can be seen from figures \ref{fig:11_b}-\ref{fig:11_c}. Moreover, when the Bond number is sufficiently large ($Bo = 10.0$), multiple cutoff wavenumbers appear (see figure \ref{fig:11_c}). The gravitational effect due to the Bond number stabilizes the flow when an external shear force is absent but behaves exactly opposite for increasing positive shear force. Here, the competing influences of these two forces cause the wave crests to move faster than the troughs, which results in rupture of the thin film. 
	
	\begin{figure}[ht!]
		\begin{center} 
			\subfigure[$\mathcal{C} = 0.01$]{\label{fig:12_a}\includegraphics*[height=4.5cm, width=5.5cm]{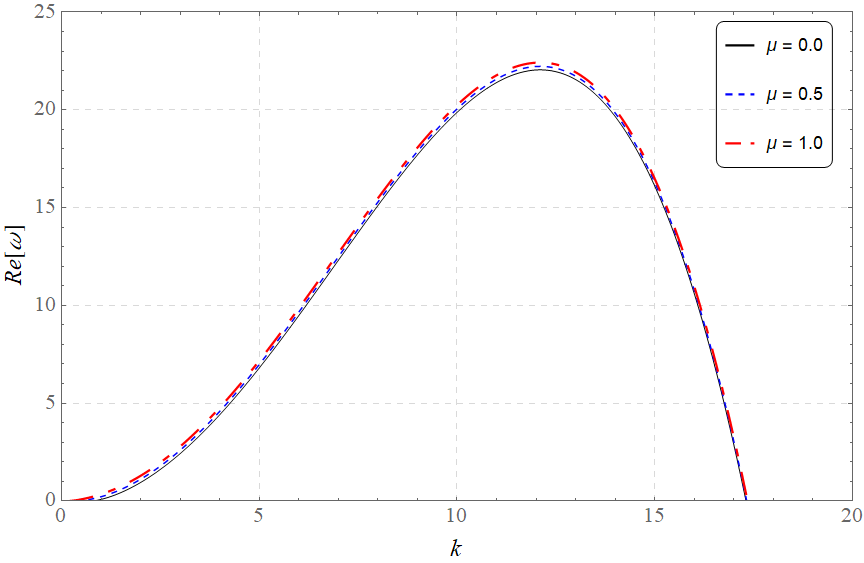}}
			\subfigure[$\mathcal{C} = 1.0$]{\label{fig:12_b}\includegraphics*[height=4.5cm, width=5.5cm]{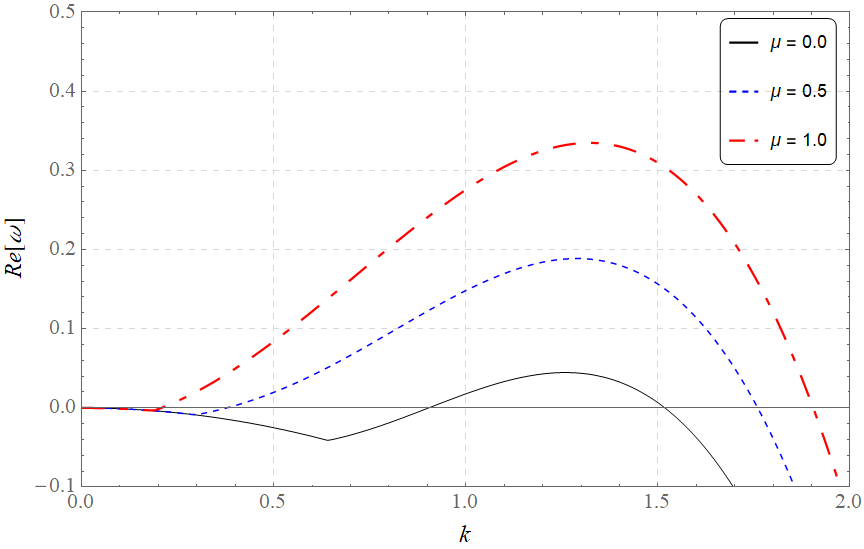}}
			\subfigure[$\mathcal{C} = 10.0$]{\label{fig:12_c}\includegraphics*[height=4.5cm, width=5.5cm]{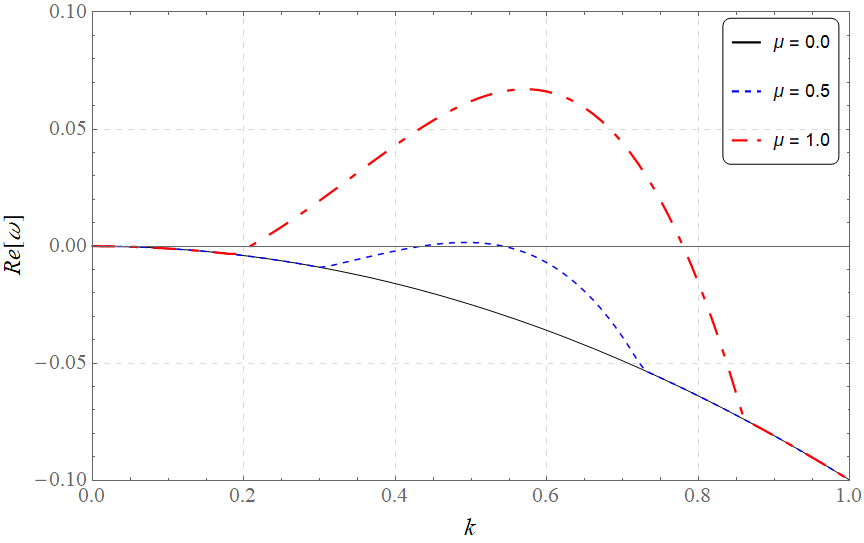}}
		\end{center} 
		\caption{ \scriptsize  The growth rate curves corresponding to the dispersion relation against wavenumbers $k$ for different external shear forces $\mu$ in case of (a) $\mathcal{C} = 0.01$, (b) $\mathcal{C} = 1.0$, and (c) $\mathcal{C} = 10.0$. The rest of the parameters are $H_b = 1.0$, $\Gamma_{0,B} = 1.0$, $\Gamma_{a,B} = 1.0$, $Bo = 10.0$, $k_a = 0.01$, $Pe_b = 10$, $Pe_{s} = 100$, $\beta_a = 1.0$, $\mathcal{A} = 1.0$, $\alpha = 0.01$, $\delta = 0.05$. } \label{fig:12}
	\end{figure}
	
	On the other hand, the effect of capillarity on the external shear force is explored in Figure \ref{fig:12_a}-\ref{fig:12_c}. It is observed from Figure \ref{fig:12_a} that when $\mathcal{C}$ is low, the magnitude perturbation growth rate is very high, and it grows with external shear force $\mu$. In the case of a unit value of Capillary force $\mathcal{C}$, the system remains stable in the long-wave regime, and when the external shear force is absent, the flow remains stable even in the finite wave zone (see Figure \ref{fig:12_b}). Moreover, for $\mathcal{C} = 10.0$ and $\mu = 0.0$, the flow is stable throughout, and even for $\mu = 0.5$, the system is almost stable, as is evident from Figure \ref{fig:12_c}. For small $\mathcal{C}$, surface tension dominates the viscous and shear-layer forces. So, at $\mathcal{C} = 0.01$, external shear force $\mu$ and surface tension compete, and from Figure \ref{fig:12_a}, it can be observed that external shear force elevates the disturbances generated due to the perturbation. However, when the capillary number increases, the viscous shear stress regulates the surface tension and external shear force. As $\mathcal{C} \uparrow $, the flow becomes stable at lower values of $\mu$ at finite wavenumbers. Moreover, the cut-off wavenumbers were also reduced for a higher capillary number.

	\begin{figure}[ht!]
		\begin{center} 
			\subfigure[$\mathcal{A} = 0.01$]{\label{fig:13_a}\includegraphics*[height=4.5cm, width=5.5cm]{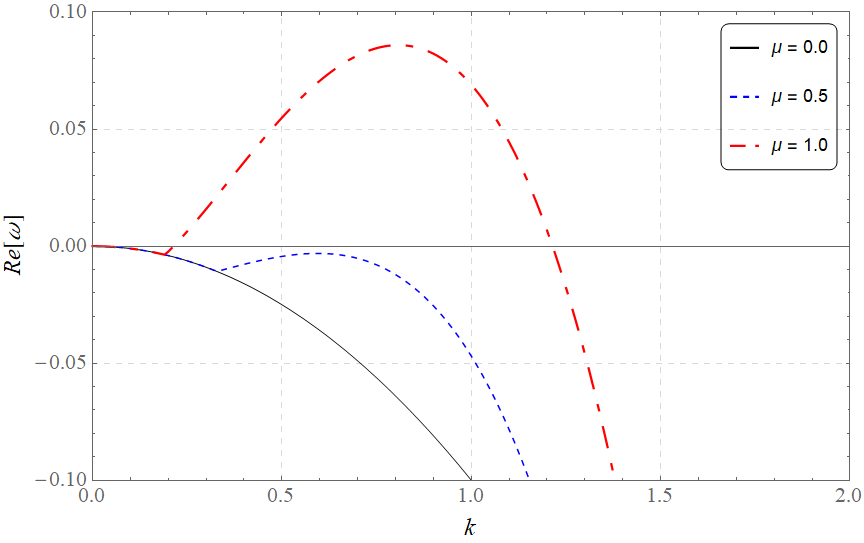}}
			\subfigure[$\mathcal{A} = 1.0$]{\label{fig:13_b}\includegraphics*[height=4.5cm, width=5.5cm]{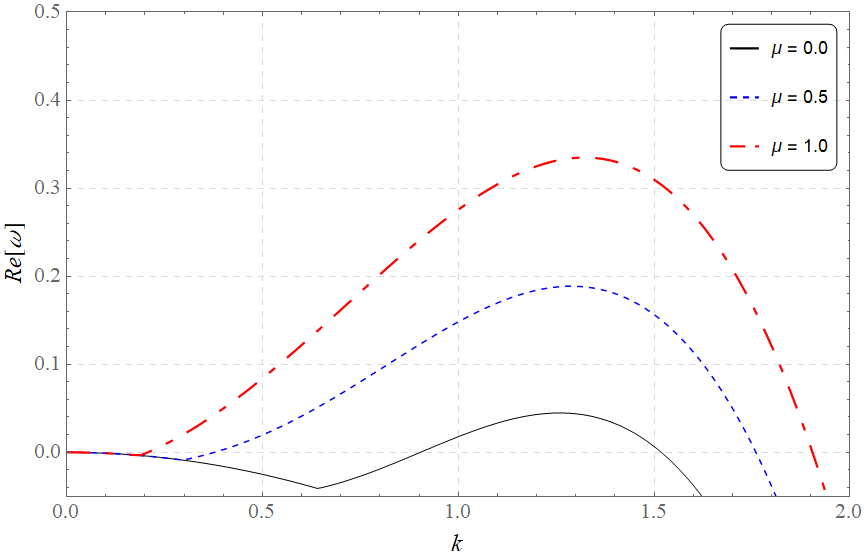}}
			\subfigure[$\mathcal{A} = 10.0$]{\label{fig:13_c}\includegraphics*[height=4.5cm, width=5.5cm]{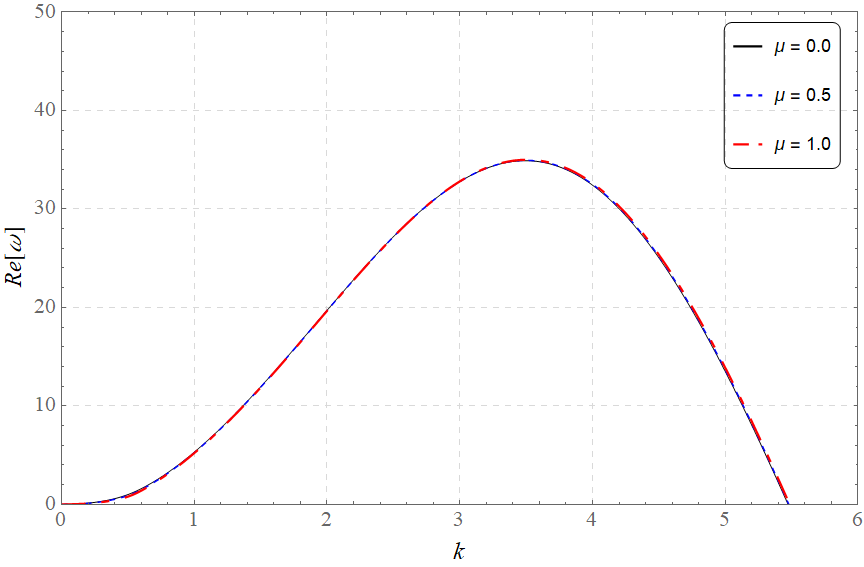}}
		\end{center} 
		\caption{ \scriptsize  The growth rate curves corresponding to the dispersion relation against wavenumbers $k$ for different external shear forces $\mu$ in case of (a) $\mathcal{A} = 0.01$, (b) $\mathcal{A} = 1.0$, and (c) $\mathcal{A} = 10.0$. The rest of the parameters are $H_b = 1.0$, $\Gamma_{0,B} = 1.0$, $\Gamma_{a,B} = 1.0$, $Bo = 10.0$, $k_a = 0.01$, $Pe_b = 10$, $Pe_{s} = 100$, $\beta_a = 1.0$, $\mathcal{C} = 1.0$, $\alpha = 0.01$, $\delta = 0.05$. } \label{fig:13}
	\end{figure}
	
	Furthermore, Figures \ref{fig:13} explores the effect of the Hamaker constant $\mathcal{A}$ on \textit{mode 3} of the linear stability analysis. It can be seen in Figure \ref{fig:13_a} that when $\mathcal{A} = 0.01$, the system is stable for all wavenumbers when external shear $\mu < 1 $. If $\mu = 1$, then only the system becomes unstable. For moderate values of $\mathcal{A}$ (read $\mathcal{A} = 1.0$ ), the flow displays an unstable nature in the finite wavenumbers regime, and when  $\mathcal{A} = 10.0$, it remains unstable for all wavenumbers (see figures \ref{fig:13_b}-\ref{fig:13_c}). Also, in each scenario, growth rates increase for increasing external shear force $\mu$. That implies that the Hamaker constant and external shear both destabilize the \textit{mode 3}. Physically, the Hamaker constant represents the van der Waals forces, which contribute to the disjoining pressure. Thus, when the flow undergoes perturbation, the thinner regions experience a more attractive force due to the van der Waals force. Also, the external shear force acts along the interface, which amplifies the disjoining pressure and helps the rupture process of the flow.

	\subsubsection{Influence of slip parameter $\delta$}
	
	\begin{figure}[ht!]
		\begin{center} 
			\subfigure[$Bo = 0.0$]{\label{fig:14_a}\includegraphics*[height=4.5cm, width=5.5cm]{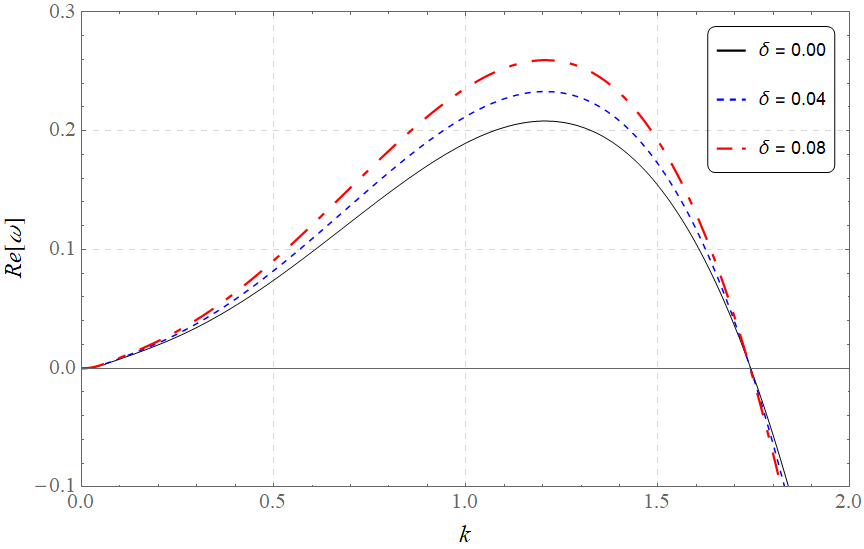}}
			\subfigure[$Bo = 1.0$]{\label{fig:14_b}\includegraphics*[height=4.5cm, width=5.5cm]{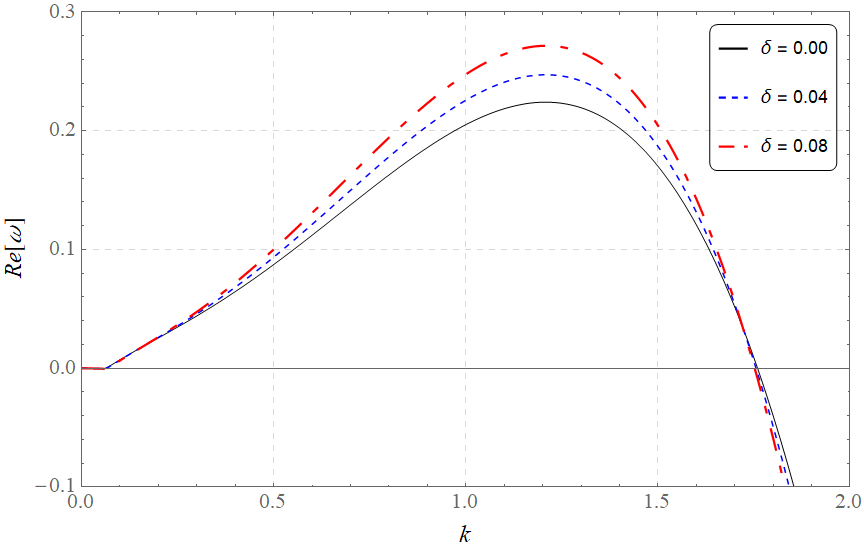}}
			\subfigure[$Bo = 10.0$]{\label{fig:14_c}\includegraphics*[height=4.5cm, width=5.5cm]{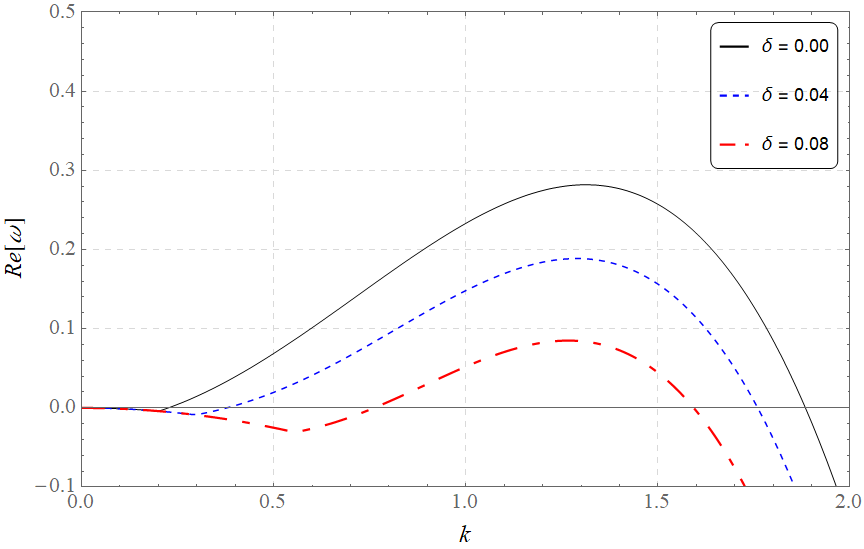}}
		\end{center} 
		\caption{ \scriptsize The growth rate curves corresponding to the dispersion relation against wavenumbers $k$ for different slip length $\delta$ in case of (a) $Bo = 0.0$, (b) $Bo = 1.0$, and (c) $Bo = 10.0$. The rest of the parameters are $H_b = 1.0$, $\Gamma_{0,B} = 1.0$, $\Gamma_{a,B} = 1.0$, $k_a = 0.01$, $Pe_b = 10$, $Pe_{s} = 100$, $\beta_a = 1.0$, $\mathcal{A} = 1.0$, $\mathcal{C} = 1.0$, $\alpha = 0.01$, $\tau = 0.5$. } \label{fig:14}
	\end{figure}
	
	The impact of the slip parameter $\delta$ on the linear stability is explored in this subsection. First of all, the dual effect of the Bond number $Bo$ and the slip parameter $\delta$ is illustrated in the figures (\ref{fig:14}). When $Bo = 0$, the slip parameter destabilizes the flow by scaling greater heights for a bigger slip length (see figure \ref{fig:14_a}). The behaviour of the slip parameter $\delta$ remains the same even when $Bo = 1$, as evident from figure \ref{fig:14_b}. However, figure \ref{fig:14_c} shows that the slip parameter shows a stabilizing nature for $Bo = 10.0$ in the finite wave regime, as in the long-wave regime, the flow remains stable. For a low Bond number, the fluid is governed by surface tension, and increasing slip length reduces the viscous drag at the bottom substrate, which increases the velocity. This amplified velocity helps to advect the soluble surfactant quickly, which generates local surfactant concentration gradients and draws more fluid towards the crests, which destabilizes the flow.
	
	\begin{figure}[ht!]
		\begin{center} 
			\subfigure[$\mathcal{C} = 0.01$]{\label{fig:15_a}\includegraphics*[height=4.5cm, width=5.5cm]{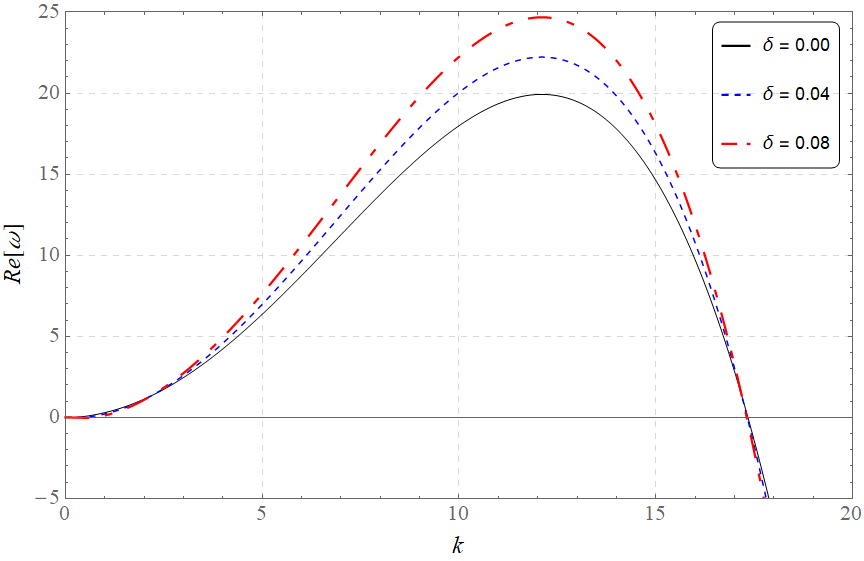}}
			\subfigure[$\mathcal{C} = 1.0$]{\label{fig:15_b}\includegraphics*[height=4.5cm, width=5.5cm]{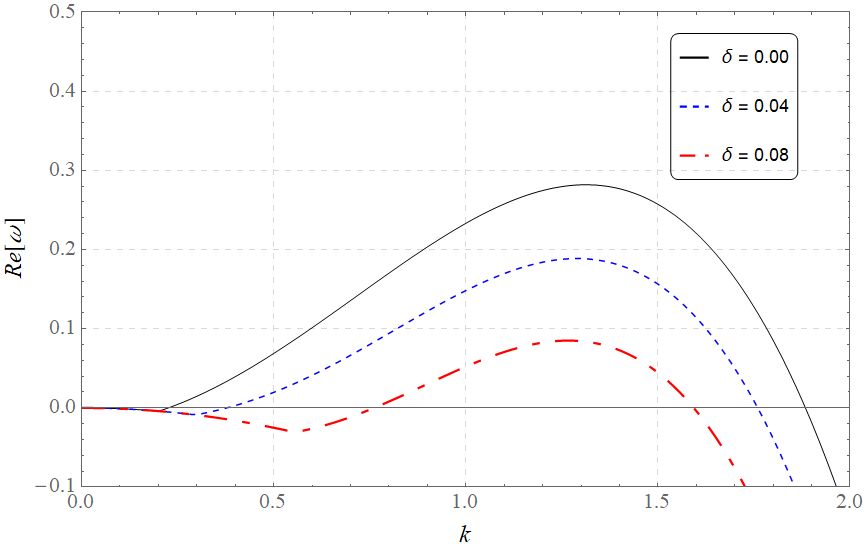}}
			\subfigure[$\mathcal{C} = 10.0$]{\label{fig:15_c}\includegraphics*[height=4.5cm, width=5.5cm]{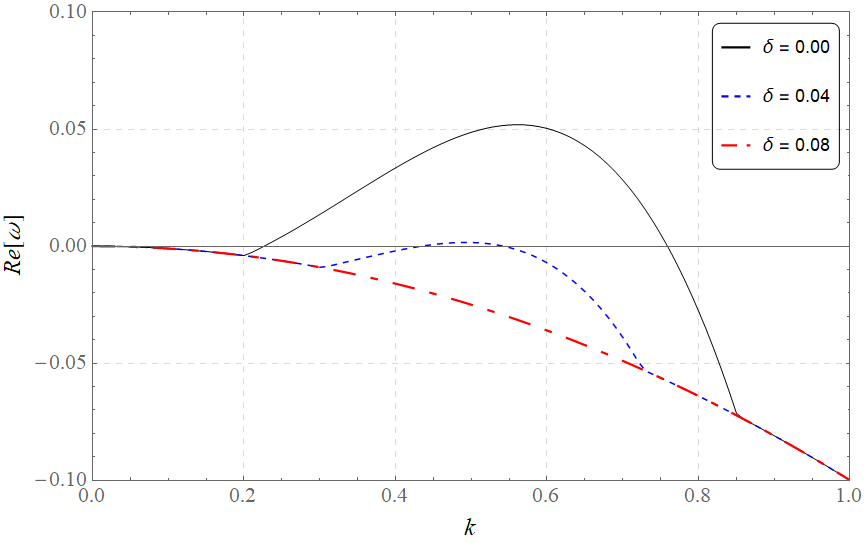}}
		\end{center} 
		\caption{ \scriptsize  The growth rate curves corresponding to the dispersion relation against wavenumbers $k$ for different external shear force $\mu$ in case of (a) $\mathcal{C} = 0.01$, (b) $\mathcal{C} = 1.0$, and (c) $\mathcal{C} = 10.0$. The rest of the parameters are $H_b = 1.0$, $\Gamma_{0,B} = 1.0$, $\Gamma_{a,B} = 1.0$, $Bo = 10.0$, $k_a = 0.01$, $Pe_b = 10$, $Pe_{s} = 100$, $\beta_a = 1.0$, $\mathcal{A} = 1.0$, $\alpha = 0.01$, $\tau = 0.5$. } \label{fig:15}
	\end{figure}
	
	Figures (\ref{fig:15}) describes the influence of capillary number $\mathcal{C} = 0.01$ and slip parameter $\delta$ on the soluble surfactant-laden fluid flow. When the capillary number is low, the slip length destabilizes the flow, as can be seen from Figure \ref{fig:15_a}. However, the slip parameter $\delta$ stabilizes the thin film flow for moderate to high capillarity. In the case of $\mathcal{C} = 1.0$, the flow, it remains stable in the long-wave regime, and cut-off wavenumbers increase for increasing slip length. Moreover, higher slip length ($\delta = 0.08$) stabilizes the flow for any wavenumbers when $\mathcal{C} = 10.0$. At low capillary numbers, surface tension dominates, and the slip parameter allows the fluid layers near the boundary to move smoothly, which amplifies the surface waves, promoting the instability. But at a high capillary number, viscous and solutal Marangoni forces dominate the other forces. In that case, increasing slip length allows more fluid flow and more circulation of surfactant between bulk and interface, which mitigates the Marangoni stress. Thus, the chance of generating the rupture due to the Marangoni stress attenuates.
	
	\begin{figure}[ht!]
		\begin{center} 
			\subfigure[$\mathcal{A} = 0.01$]{\label{fig:16_a}\includegraphics*[height=4.5cm, width=5.5cm]{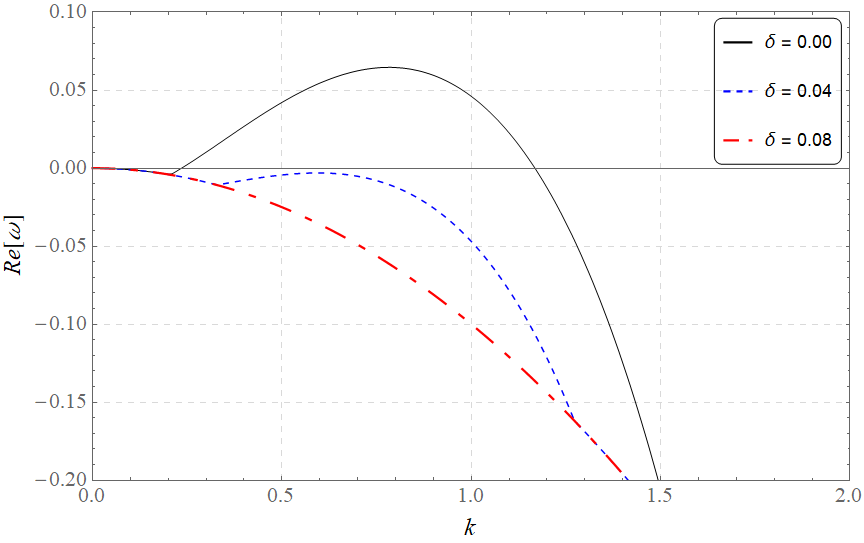}}
			\subfigure[$\mathcal{A} = 1.0$]{\label{fig:16_b}\includegraphics*[height=4.5cm, width=5.5cm]{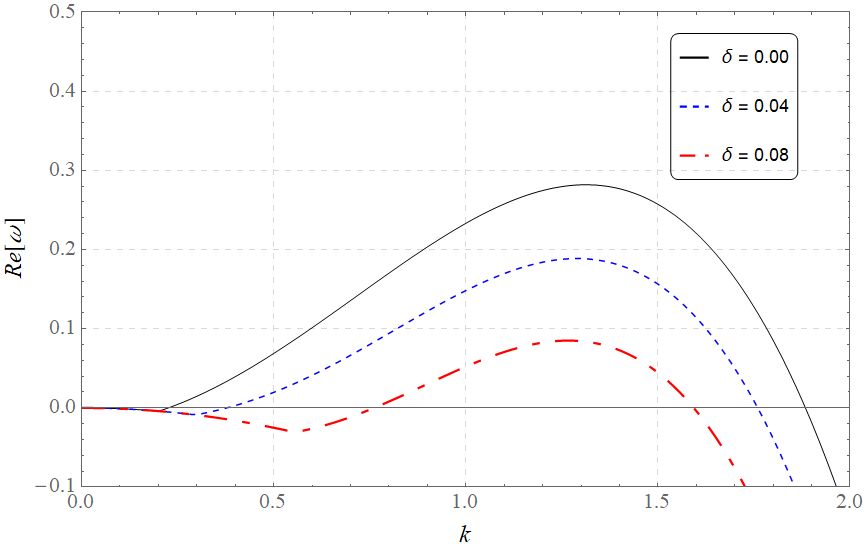}}
			\subfigure[$\mathcal{A} = 10.0$]{\label{fig:16_c}\includegraphics*[height=4.5cm, width=5.5cm]{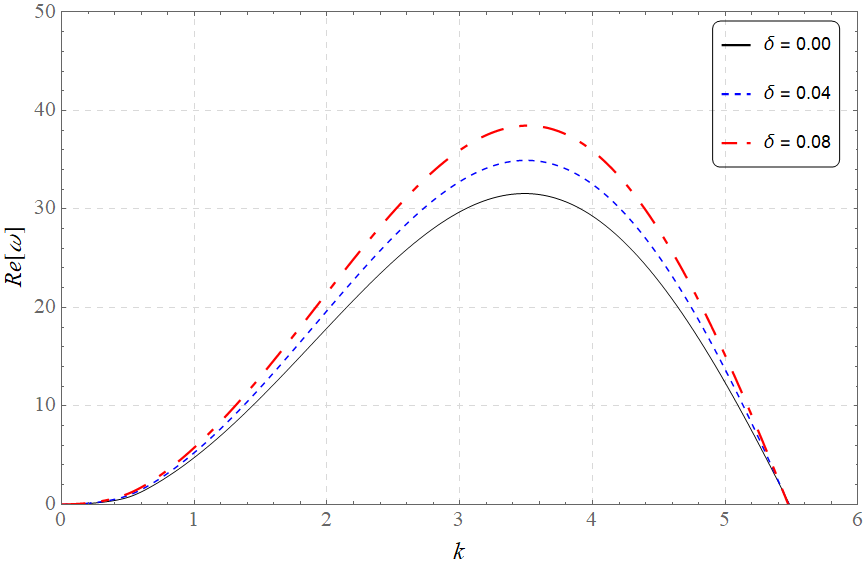}}
		\end{center} 
		\caption{ \scriptsize  The growth rate curves corresponding to the dispersion relation against wavenumbers $k$ for different external shear forces $\mu$ in the case of (a) $\mathcal{A} = 0.01$, (b) $\mathcal{A} = 1.0$, and (c) $\mathcal{A} = 10.0$. The rest of the parameters are $H_b = 1.0$, $\Gamma_{0,B} = 1.0$, $\Gamma_{a,B} = 1.0$, $Bo = 10.0$, $k_a = 0.01$, $Pe_b = 10$, $Pe_{s} = 100$, $\beta_a = 1.0$, $\mathcal{C} = 1.0$, $\alpha = 0.01$, $\tau = 0.5$. } \label{fig:16}
	\end{figure}
	
	The binary effect of the slip parameter $\delta$ and Hamaker constant $\mathcal{A}$ are studied in Figure (\ref{fig:16}). The slip parameter $\delta$ has a stabilizing effect on the flow when the Hamaker constant is small. In fact, Figure \ref{fig:16_a} shows that for $\mathcal{A} = 0.01$, the system is unstable when the no-slip condition is imposed at the bottom substrate and becomes stable when the slip parameter $\delta > 0$. For a moderate value of $\mathcal{A}$($ = 1.0$), the system becomes stable for increasing slip length in the finite wavenumbers zone, and in the long-wave regime, the flow is stable irrespective of the slip length (see figure \ref{fig:16_b}). Moreover, the cutoff wavenumbers amplify for increasing slip length $\delta$. On the contrary, the growth rates increase for increasing slip length when the Hamaker constant $\mathcal{A} = 10.0$, implying the destabilizing nature of slip length $\delta$ for higher van der Waals force (see figure \ref{fig:16_c}). When $\mathcal{A} = 0.01$, the van der Waals force is weak; a slippery substrate helps the fluid flow easily into the thinning regions and reduces the growth of disturbances and stabilizes the thin film. But when $\mathcal{A} = 10.0$, the attractive van der Waals forces pull the fluid out of the thinning region, and a higher slip length drives the fluid more efficiently; it increases the chance of destabilizing the flow and rupturing the film.

	\section{SUMMARY AND CONCLUSION} \label{sec:SC}
	
	Spreading and rupture dynamics of soluble surfactant-laden fluid over a slippery bottom in the presence of external shear are investigated in the article. The lubrication approximation method is used to derive the evolution equation of the film thickness and interface surfactant concentration. Moreover, the rapid vertical diffusion is assumed to eliminate the fingering instability, and then the cross-sectionally averaged bulk surfactant concentration is considered. The equations are particularly suitable to describe the flow nature during the surfactant distribution and evanescent period when sorption kinetics is in effect. This evanescent period is quantified through $\mathcal{O}(1/k_a)$, where $k_a$ denotes the time taken by the sorption kinetics to take place. The shear stress promotes horizontal and vertical velocities. This vertical velocity-induced flow causes the fluid thickness to increase, and the local bulk surfactant concentration is diluted. This causes the flow to experience a squeezing action, which results in a welling up of fluid. This article focused on the influence of external shear force and slip length on this welling up of the fluid. Based on $k_a$ time, the study considers two different scenarios: when $k_a \gg 1$, rapid sorption kinetics happen, i.e., the sorption reaction happens very fast, and the whole system comes to equilibrium. On the other hand, when $k_a \ll 1$, slow sorption kinetics occur; the sorption reaction takes time to reach equilibrium. Again, in the case of rapid sorption kinetics, two types of surfactant distribution, planar strip and planar front, are considered. These are basically based on different supply rates of the surfactant with $\propto t^{1/3}$ for the strip and $\propto t^{1/2}$ for the front. In the case of strip-type distribution of surfactant for rapid sorption kinetics, the slippery bottom helps the capillary ridge move faster, whereas the external shear force $\tau$ acts as an auxiliary force along with the Marangoni stress, which amplifies the thinning of the film. The streamfunction plots show that a distinct recirculation zone is created next to the surfactant distribution, and the slippery substrate helps more fluid to flow by lowering viscous drag at the bottom and the shock front to gain more volume and advance horizontally. In contrast, the front-type distribution of rapid sorption kinetics generates a hump at the leading edge of the monomer followed by a depressed thinning region. The slippery bottom $\delta$ reduces the viscous dissipation at the boundary, resulting in a taller hump and an even steeper depression at the trailing end behind the hump. The external shear force $\mu$ pushes the hump downstream and reduces the fluid shock height. Furthermore, the front-type deposition of surfactant in the case of slow sorption kinetics reveals that bulk and interface surfactant concentration go through an adsorption-desorption transient stage, which eventually reduces the surface tension gradient at the interface, i.e., the Marangoni stress diminishes, resulting in pulse-type behaviour in the film thickness. The external shear force $\mu$ acts along the interface of the fluid and reduces the pulse height systematically, but the slip substrate at the bottom allows more fluid to flow and the pulse to gain more volume. 
	
	The linear stability analysis using the rupture dynamics mechanism is explored from the evolution equation obtained earlier. The van der Waals forces are considered to be one of the major factors behind the rupture of the thin film. The results clearly state that the Bond number $Bo$ destabilizes the flow in the presence of external shear in the finite wave regime. However, the flow remains stable in the long-wave zone for a positive Bond number. In contrast, capillarity stabilizes the flow, whereas the Hamaker constant destabilizes it. In all scenarios, the external shear force $\mu$ destabilizes the flow. Conversely, the slip parameter $\delta$ has a binary effect on the linear stability of the rupture dynamics of the flow. For a low Bond number $Bo$, the slip parameter destabilizes the flow, but for high Bo, this trend reverses. The same characteristics can be observed for capillary number $\mathcal{C}$. The slip parameter stabilizes the flow for low to moderate values of the Hamaker constant $\mathcal{A}$, but destabilizes for high $\mathcal{A}$. 
	
	\begin{acknowledgments}
		The first author would like to acknowledge the help from Dr. Prashant Kumar Pandey (Department of Applied Science, Amity University, Ranchi, Jharkhand, India) during the preparation of this manuscript. Also, he acknowledged the financial support from SERB, Department of Science and Technology, Government of India, through the CRG project, Award No. CRG/2022/006698.
	\end{acknowledgments}
	
	\section{Appendix}\label{sec:apndx}
	
	\begin{align}
		a_1 = & \displaystyle \nonumber \Bigg[-\frac{1}{2} \mathrm{i} Bo H_b^2 k-k_a-\frac{k^2}{Pe_b}-\frac{k^2}{Pe_{s}}-k_a R_a \Gamma_{0,B}+\frac{k_a R_a \beta_a (-1+\Gamma_{a,B})}{H_b}-\mathrm{i} Bo H_b k (H_b+\delta)\\
		& \displaystyle \nonumber -\frac{3 k^2 \left(1+\left(1+\frac{1}{\alpha }\right)^{1/3}\right) (1+\alpha ) \Gamma_{a,B} (H_b+\delta )}{\left(1+\Gamma_{a,B}+\left(1+\frac{1}{\alpha }\right)^{1/3} \Gamma_{a,B}\right)^4} -\frac{1}{6} \mathrm{i} Bo H_b k (2 H_b+3 \delta ) \displaystyle \nonumber-\frac{k^2 \left(-3 A+C H_b^4 k^2\right) (2 H_b+3 \delta )}{6 H_b^2}-3 \mathrm{i} k (H_b+\delta ) \tau \Bigg]
	\end{align}

	\begin{align}
		a_2 = & \nonumber \Bigg[-\frac{k^2 k_a}{Pe_b}-\frac{k^4}{Pe_b Pe_{s}}-\frac{k^2 k_a R_a \Gamma_{0,B}}{Pe_b} +\frac{k_a^2 R_a \beta_a (-1+\Gamma_{a,B})}{H_b}+\frac{k^2 k_a R_a \beta_a (-1+\Gamma_{a,B})}{H_b Pe_{s}}\\
		& \nonumber +\frac{k_a^2 R_a^2 \beta_a \Gamma_{0,B} (-1+ \Gamma_{a,B})}{H_b}+\frac{k_a^2 R_a \beta_a (1+R_a \Gamma_{0,B}) (-1+\Gamma_{a,B})}{H_b}-\mathrm{i} Bo H_b k k_a (H_b+\delta )\\
		& \nonumber -\frac{\mathrm{i} Bo H_b k^3 (H_b+\delta )}{Pe_b}-\frac{\mathrm{i} Bo H_b k^3 (H_b+\delta)}{Pe_{s}}-\mathrm{i} Bo H_b k k_a R_a \Gamma_{0,B} (H_b+\delta )\\
		& \nonumber +\mathrm{i} Bo k k_a R_a \beta_a (-1+\Gamma_{a,B}) (H_b+\delta )-\frac{3 k^4 \left(1+\left(1+\frac{1}{\alpha }\right)^{1/3}\right) (1+\alpha ) \Gamma_{a,B} (H_b+\delta )}{Pe_b \left(1+\Gamma_{a,B}+\left(1+\frac{1}{\alpha }\right)^{1/3} \Gamma_{a,B}\right)^4}\\
		& \nonumber +\frac{3 k^2 k_a R_a \left(1+\left(1+\frac{1}{\alpha }\right)^{1/3}\right) (1+\alpha ) \beta_a (-1+\Gamma_{a,B}) \Gamma_{a,B} (H_b+\delta )}{H_b \left(1+\Gamma_{a,B}+\left(1+\frac{1}{\alpha }\right)^{1/3} \Gamma_{a,B}\right)^4}\\
		& \nonumber -\frac{3 \mathrm{i} Bo H_b k^3 \left(1+\left(1+\frac{1}{\alpha }\right)^{1/3}\right) (1+\alpha ) \Gamma_{a,B} (H_b+\delta )^2}{\left(1+\Gamma_{a,B}+\left(1+\frac{1}{\alpha }\right)^{1/3} \Gamma_{a,B}\right)^4}-\frac{1}{6} \mathrm{i} Bo H_b k k_a (2 H_b+3 \delta )\\
		& \nonumber -\frac{k^2 \left(-3 A+C H_b^4 k^2\right) k_a (2 H_b+3 \delta )}{6 H_b^2}-\frac{k^4 \left(-3 A+C H_b^4 k^2\right) (2 H_b+3 \delta )}{6 H_b^2 Pe_b}-\frac{\mathrm{i} Bo H_b k^3 (2 H_b+3 \delta )}{6 Pe_{s}}\\
		& \nonumber -\frac{k^4 \left(-3 A+C H_b^4 k^2\right) (2 H_b+3 \delta )}{6 H_b^2 Pe_{s}}-\frac{1}{6} \mathrm{i} Bo H_b k k_a R_a \Gamma_{0,B} (2 H_b+3 \delta )\\
		& \nonumber -\frac{k^2 \left(-3 A+C H_b^4 k^2\right) k_a R_a \Gamma_{0,B} (2 H_b+3 \delta )}{6 H_b^2}+\frac{k^2 \left(-3 A+C H_b^4 k^2\right) k_a R_a \beta_a (-1+ \Gamma_{a,B}) (2 H_b+3 \delta )}{6 H_b^3}\\
		& \nonumber +\frac{1}{6} Bo^2 H_b^2 k^2 (H_b+\delta ) (2 H_b+3 \delta )-\frac{\mathrm{i} Bo H_b k^3 \left(1+\left(1+\frac{1}{\alpha }\right)^{1/3}\right) (1+\alpha ) \Gamma_{a,B} (H_b+\delta ) (2 H_b+3 \delta )}{2 \left(1+\Gamma_{a,B}+\left(1+\frac{1}{\alpha }\right)^{1/3} \Gamma_{a,B}\right)^4}\\
		& \nonumber -\frac{k^4 \left(-3 A+C H_b^4 k^2\right) \left(1+\left(1+\frac{1}{\alpha }\right)^{1/3}\right) (1+\alpha ) \Gamma_{a,B} (H_b+\delta ) (2 H_b+3 \delta )}{2 H_b^2 \left(1+\Gamma_{a,B}+\left(1+\frac{1}{\alpha }\right)^{1/3} \Gamma_{a,B}\right)^4}
	\end{align}
	
	\begin{align}
		& \nonumber -\frac{\mathrm{i} Bo k^3 \left(-3 A+C H_b^4 k^2\right) (2 H_b+3 \delta )^2}{36 H_b}-2 \mathrm{i} k k_a (H_b+\delta ) \tau -\frac{\mathrm{i} k^3 (H_b+\delta ) \tau }{Pe_b}-\frac{2 \mathrm{i} k^3 (H_b+\delta ) \tau }{Pe_{s}}\\
		& \nonumber -2 \mathrm{i} k k_a R_a \Gamma_{0,B} (H_b+\delta ) \tau +\frac{\mathrm{i} k k_a R_a \beta_a (-1+\Gamma_{a,B}) (H_b+\delta ) \tau }{H_b}+Bo H_b k^2 (H_b+\delta )^2 \tau\\
		& \nonumber -\frac{6 \mathrm{i} k^3 \left(1+\left(1+\frac{1}{\alpha }\right)^{1/3}\right) (1+\alpha ) \Gamma_{a,B} (H_b+\delta )^2 \tau }{\left(1+\Gamma_{a,B}+\left(1+\frac{1}{\alpha }\right)^{1/3} \Gamma_{a,B}\right)^4}+\frac{1}{6} Bo H_b k^2 (H_b+\delta ) (2 H_b+3 \delta ) \tau \\
		& \nonumber -\frac{\mathrm{i} k^3 \left(-3 A+C H_b^4 k^2\right) (H_b+\delta ) (2 H_b+3 \delta ) \tau }{6 H_b^2}+k^2 (H_b+\delta )^2 \tau ^2\\   
		& \nonumber +\frac{3 k^3 \left(1+\left(1+\frac{1}{\alpha }\right)^{1/3}\right) (1+\alpha ) \Gamma_{a,B} (H_b+2 \delta ) \left(2 \mathrm{i} Bo H_b^3-3 A k+C H_b^4 k^3+2 \mathrm{i} H_b^2 \tau \right)}{4 H_b \left(1+ \Gamma_{a,B}+\left(1+\frac{1}{\alpha }\right)^{1/3} \Gamma_{a,B}\right)^4}\\
		& \nonumber -\frac{\mathrm{i} k^3 \left(\frac{Bo H_b^2}{2}+(H_b+\delta ) \tau \right)}{Pe_b}+\frac{\mathrm{i} k k_a R_a \beta_a (-1+\Gamma_{a,B}) \left(\frac{Bo H_b^2}{2}+(H_b+\delta) \tau \right)}{H_b}\\
		& \nonumber +Bo H_b k^2 (H_b+\delta ) \left(\frac{Bo H_b^2}{2}+(H_b+\delta ) \tau \right)+\frac{1}{12} Bo H_b k^2 (2 H_b+3 \delta ) \left(Bo H_b^2+2 (H_b+\delta ) \tau \right)\\
		& \nonumber -\frac{\mathrm{i} k^3 \left(-3 A+C H_b^4 k^2\right) (2 H_b+3 \delta ) \left(Bo H_b^2+2 (H_b+\delta ) \tau \right)}{12 H_b^2}+k^2 (H_b+\delta ) \tau  \left(Bo H_b^2+2 (H_b+\delta ) \tau \right)\Bigg]
	\end{align}
	
	\begin{align}
		a_3 = & \nonumber \Bigg[ -\frac{\mathrm{i} Bo H_b k^3 k_a (H_b+\delta )}{Pe_b}-\frac{\mathrm{i} Bo H_b k^5 (H_b+\delta )}{Pe_b Pe_{s}}-\frac{\mathrm{i} Bo H_b k^3 k_a R_a \Gamma_{0,B} (H_b+\delta )}{Pe_b}\\
		& \nonumber +\mathrm{i} Bo k k_a^2 R_a \beta_a (-1+\Gamma_{a,B}) (H_b+\delta )+\frac{\mathrm{i} Bo k^3 k_a R_a \beta_a (-1+\Gamma_{a,B}) (H_b+\delta )}{Pe_{s}}\\
		& \nonumber +\mathrm{i} Bo k k_a^2 R_a^2 \beta_a \Gamma_{0,B} (-1+\Gamma_{a,B}) (H_b+\delta )+\mathrm{i} Bo k k_a^2 R_a \beta_a (1+R_a \Gamma_{0,B}) (-1+\Gamma_{a,B}) (H_b+\delta )\\
		& \nonumber -\frac{3 \mathrm{i} Bo H_b k^5 \left(1+\left(1+\frac{1}{\alpha }\right)^{1/3}\right) (1+\alpha ) \Gamma_{a,B} (H_b+\delta )^2}{Pe_b \left(1+ \Gamma_{a,B}+\left(1+\frac{1}{\alpha }\right)^{1/3} \Gamma_{a,B}\right)^4}\\
		& \nonumber +\frac{3 \mathrm{i} Bo k^3 k_a R_a \left(1+\left(1+\frac{1}{\alpha }\right)^{1/3}\right) (1+\alpha ) \beta_a (-1+\Gamma_{a,B}) \Gamma_{a,B} (H_b+\delta )^2}{\left(1+\Gamma_{a,B}+\left(1+\frac{1}{\alpha }\right)^{1/3} \Gamma_{a,B}\right)^4}\\
		& \nonumber +\frac{3 k^2 k_a^2 R_a \left(1+\left(1+\frac{1}{\alpha }\right)^{1/3}\right) (1+\alpha ) \beta_a (-1+\Gamma_{a,B}) (R_a \Gamma_{0,B} (-1+\Gamma_{a,B})+\Gamma_{a,B}) (H_b+2 \delta )}{2 H_b \left(1+ \Gamma_{a,B}+\left(1+\frac{1}{\alpha }\right)^{1/3} \Gamma_{a,B}\right)^4}\\
		& \nonumber -\frac{k^4 \left(-3 A+C H_b^4 k^2\right) k_a (2 H_b+3 \delta )}{6 H_b^2 Pe_b}-\frac{k^6 \left(-3 A+C H_b^4 k^2\right) (2 H_b+3 \delta )}{6 H_b^2 Pe_b Pe_{s}}\\
		& \nonumber -\frac{k^4 \left(-3 A+C H_b^4 k^2\right) k_a R_a \Gamma_{0,B} (2 H_b+3 \delta )}{6 H_b^2 Pe_b}+\frac{k^2 \left(-3 A+C H_b^4 k^2\right) k_a^2 R_a \beta_a (-1+\Gamma_{a,B}) (2 H_b+3 \delta )}{6 H_b^3}\\
		& \nonumber +\frac{k^4 \left(-3 A+C H_b^4 k^2\right) k_a R_a \beta_a (-1+\Gamma_{a,B}) (2 H_b+3 \delta )}{6 H_b^3 Pe_{s}}\\
		& \nonumber +\frac{k^2 \left(-3 A+C H_b^4 k^2\right) k_a^2 R_a^2 \beta_a \Gamma_{0,B} (-1+\Gamma_{a,B}) (2 H_b+3 \delta )}{6 H_b^3}\\
		& \nonumber +\frac{k^2 \left(-3 A+C H_b^4 k^2\right) k_a^2 R_a \beta_a (1+R_a \Gamma_{0,B}) (-1+ \Gamma_{a,B}) (2 H_b+3 \delta )}{6 H_b^3}+\frac{1}{6} Bo^2 H_b^2 k^2 k_a (H_b+\delta ) (2 H_b+3 \delta )
	\end{align}
	
	\begin{align}
		& \nonumber +\frac{Bo^2 H_b^2 k^4 (H_b+\delta ) (2 H_b+3 \delta )}{6 Pe_{s}}+\frac{1}{6} Bo^2 H_b^2 k^2 k_a R_a \Gamma_{0,B} (H_b+\delta ) (2 H_b+3 \delta )\\
		& \nonumber -\frac{k^6 \left(-3 A+C H_b^4 k^2\right) \left(1+\left(1+\frac{1}{\alpha }\right)^{1/3}\right) (1+\alpha ) \Gamma_{a,B} (H_b+\delta ) (2 H_b+3 \delta )}{2 H_b^2 Pe_b \left(1+\Gamma_{a,B}+\left(1+\frac{1}{\alpha }\right)^{1/3} \Gamma_{a,B}\right)^4}\\
		& \nonumber +\frac{k^4 \left(-3 A+C H_b^4 k^2\right) k_a R_a \left(1+\left(1+\frac{1}{\alpha }\right)^{1/3}\right) (1+\alpha ) \beta_a (-1+\Gamma_{a,B}) \Gamma_{a,B} (H_b+\delta ) (2 H_b+3 \delta )}{2 H_b^3 \left(1+ \Gamma_{a,B}+\left(1+\frac{1}{\alpha }\right)^{1/3} \Gamma_{a,B}\right)^4}\\
		& \nonumber +\frac{Bo^2 H_b^2 k^4 \left(1+\left(1+\frac{1}{\alpha }\right)^{1/3}\right) (1+\alpha ) \Gamma_{a,B} (H_b+\delta )^2 (2 H_b+3 \delta )}{2 \left(1+\Gamma_{a,B}+\left(1+\frac{1}{\alpha }\right)^{1/3} \Gamma_{a,B}\right)^4}\\
		& \nonumber -\frac{\mathrm{i} Bo k^3 \left(-3 A+C H_b^4 k^2\right) k_a (2 H_b+3 \delta )^2}{36 H_b}-\frac{\mathrm{i} Bo k^5 \left(-3 A+C H_b^4 k^2\right) (2 H_b+3 \delta )^2}{36 H_b Pe_{s}}\\
		& \nonumber -\frac{\mathrm{i} Bo k^3 \left(-3 A+C H_b^4 k^2\right) k_a R_a \Gamma_{0,B} (2 H_b+3 \delta )^2}{36 H_b}\\
		& \nonumber -\frac{\mathrm{i} Bo k^5 \left(-3 A+C H_b^4 k^2\right) \left(1+\left(1+\frac{1}{\alpha }\right)^{1/3}\right) (1+\alpha ) \Gamma_{a,B} (H_b+\delta ) (2 H_b+3 \delta )^2}{12 H_b \left(1+\Gamma_{a,B}+\left(1+\frac{1}{\alpha }\right)^{1/3} \Gamma_{a,B}\right)^4}\\
		& \nonumber -\frac{\mathrm{i} k^3 k_a (H_b+\delta ) \tau }{Pe_b}-\frac{\mathrm{i} k^5 (H_b+\delta ) \tau }{Pe_b Pe_{s}}-\frac{\mathrm{i} k^3 k_a R_a \Gamma_{0,B} (H_b+\delta ) \tau }{Pe_b}+\frac{\mathrm{i} k k_a^2 R_a \beta_a (-1+\Gamma_{a,B}) (H_b+\delta ) \tau }{H_b}\\
		& \nonumber +\frac{\mathrm{i} k^3 k_a R_a \beta_a (-1+\Gamma_{a,B}) (H_b+\delta ) \tau }{H_b Pe_{s}}+\frac{\mathrm{i} k k_a^2 R_a^2 \beta_a \Gamma_{0,B} (-1+\Gamma_{a,B}) (H_b+\delta ) \tau }{H_b}\\
		& \nonumber +\frac{\mathrm{i} k k_a^2 R_a \beta_a (1+R_a \Gamma_{0,B}) (-1+ \Gamma_{a,B}) (H_b+\delta ) \tau }{H_b}+Bo H_b k^2 k_a (H_b+\delta )^2 \tau +\frac{Bo H_b k^4 (H_b+\delta )^2 \tau }{Pe_{s}}\\
		& \nonumber +Bo H_b k^2 k_a R_a \Gamma_{0,B} (H_b+\delta )^2 \tau -\frac{3 \mathrm{i} k^5 \left(1+\left(1+\frac{1}{\alpha }\right)^{1/3}\right) (1+\alpha ) \Gamma_{a,B} (H_b+\delta )^2 \tau }{Pe_b \left(1+\Gamma_{a,B}+\left(1+\frac{1}{\alpha }\right)^{1/3} \Gamma_{a,B}\right)^4}\\
		& \nonumber +\frac{3 \mathrm{i} k^3 k_a R_a \left(1+\left(1+\frac{1}{\alpha }\right)^{1/3}\right) (1+\alpha ) \beta_a (-1+\Gamma_{a,B}) \Gamma_{a,B} (H_b+\delta )^2 \tau }{H_b \left(1+\Gamma_{a,B}+\left(1+\frac{1}{\alpha }\right)^{1/3} \Gamma_{a,B}\right)^4}\\
		& \nonumber +\frac{3 Bo H_b k^4 \left(1+\left(1+\frac{1}{\alpha }\right)^{1/3}\right) (1+\alpha ) \Gamma_{a,B} (H_b+\delta )^3 \tau }{\left(1+\Gamma_{a,B}+\left(1+\frac{1}{\alpha }\right)^{1/3} \Gamma_{a,B}\right)^4}+\frac{1}{6} Bo H_b k^2 k_a (H_b+\delta ) (2 H_b+3 \delta ) \tau\\
		& \nonumber -\frac{\mathrm{i} k^3 \left(-3 A+C H_b^4 k^2\right) k_a (H_b+\delta ) (2 H_b+3 \delta ) \tau }{6 H_b^2}+\frac{Bo H_b k^4 (H_b+\delta ) (2 H_b+3 \delta ) \tau }{6 Pe_{s}}\\
		& \nonumber -\frac{\mathrm{i} k^5 \left(-3 A+C H_b^4 k^2\right) (H_b+\delta ) (2 H_b+3 \delta ) \tau }{6 H_b^2 Pe_{s}}+\frac{1}{6} Bo H_b k^2 k_a R_a \Gamma_{0,B} (H_b+\delta ) (2 H_b+3 \delta ) \tau\\
		& \nonumber -\frac{\mathrm{i} k^3 \left(-3 A+C H_b^4 k^2\right) k_a R_a \Gamma_{0,B} (H_b+\delta ) (2 H_b+3 \delta ) \tau }{6 H_b^2}\\
		& \nonumber +\frac{Bo H_b k^4 \left(1+\left(1+\frac{1}{\alpha }\right)^{1/3}\right) (1+\alpha ) \Gamma_{a,B} (H_b+\delta )^2 (2 H_b+3 \delta ) \tau }{2 \left(1+\Gamma_{a,B}+\left(1+\frac{1}{\alpha }\right)^{1/3} \Gamma_{a,B}\right)^4}\\
		& \nonumber -\frac{\mathrm{i} k^5 \left(-3 A+C H_b^4 k^2\right) \left(1+\left(1+\frac{1}{\alpha }\right)^{1/3}\right) (1+\alpha ) \Gamma_{a,B} (H_b+\delta )^2 (2 H_b+3 \delta ) \tau }{2 H_b^2 \left(1+\Gamma_{a,B}+\left(1+\frac{1}{\alpha }\right)^{1/3} \Gamma_{a,B}\right)^4}+k^2 k_a (H_b+\delta )^2 \tau ^2
	\end{align}
	
	\begin{align}
		& \nonumber +\frac{k^4 (H_b+\delta )^2 \tau ^2}{Pe_{s}}+k^2 k_a R_a \Gamma_{0,B} (H_b+\delta )^2 \tau ^2\\
		& \nonumber +\frac{3 k^4 \left(1+\left(1+\frac{1}{\alpha }\right)^{1/3}\right) (1+\alpha ) \Gamma_{a,B} (H_b+\delta )^3 \tau ^2}{\left(1+\Gamma_{a,B}+\left(1+\frac{1}{\alpha }\right)^{1/3} \Gamma_{a,B}\right)^4}\\
		& \nonumber +\frac{3 k^5 \left(1+\left(1+\frac{1}{\alpha }\right)^{1/3}\right) (1+\alpha ) \Gamma_{a,B} (H_b+2 \delta ) \left(2 \mathrm{i} Bo H_b^3-3 A k+C H_b^4 k^3+2 \mathrm{i} H_b^2 \tau \right)}{4 H_b Pe_b \left(1+\Gamma_{a,B}+\left(1+\frac{1}{\alpha }\right)^{1/3}  \Gamma_{a,B}\right)^4}\\
		& \nonumber -\frac{3 k^3 k_a R_a \left(1+\left(1+\frac{1}{\alpha }\right)^{1/3}\right) (1+\alpha ) \beta_a (-1+ \Gamma_{a,B}) \Gamma_{a,B} (H_b+2 \delta ) \left(2 \mathrm{i} Bo H_b^3-3 A k+C H_b^4 k^3+2 \mathrm{i} H_b^2 \tau \right)}{4 H_b^2 \left(1+\Gamma_{a,B}+\left(1+\frac{1}{\alpha }\right)^{1/3} \Gamma_{a,B}\right)^4}\\
		& \nonumber +\frac{\mathrm{i} Bo k^4 \left(1+\left(1+\frac{1}{\alpha }\right)^{1/3}\right) (1+\alpha ) \Gamma_{a,B} (H_b+2 \delta ) (2 H_b+3 \delta ) \left(2 \mathrm{i} Bo H_b^3-3 A k+C H_b^4 k^3+2 \mathrm{i} H_b^2 \tau \right)}{8 \left(1+\Gamma_{a,B}+\left(1+\frac{1}{\alpha }\right)^{1/3} \Gamma_{a,B}\right)^4}\\
		& \nonumber +\frac{3 \mathrm{i} k^4 \left(1+\left(1+\frac{1}{\alpha }\right)^{1/3}\right) (1+\alpha ) \Gamma_{a,B} (H_b+\delta ) (H_b+2 \delta ) \tau  \left(2 \mathrm{i} Bo H_b^3-3 A k+C H_b^4 k^3+2 \mathrm{i} H_b^2 \tau \right)}{4 H_b \left(1+\Gamma_{a,B}+\left(1+\frac{1}{\alpha }\right)^{1/3}  \Gamma_{a,B}\right)^4}\\
		& \nonumber +\frac{Bo H_b k^4 (H_b+\delta ) \left(\frac{Bo H_b^2}{2}+(H_b+\delta ) \tau \right)}{Pe_b}-Bo k^2 k_a R_a \beta_a (-1+\Gamma_{a,B}) (H_b+\delta ) \left(\frac{Bo H_b^2}{2}+(H_b+\delta ) \tau \right)\\
		& \nonumber+\frac{k^4 (H_b+\delta ) \tau  \left(\frac{Bo H_b^2}{2}+(H_b+\delta ) \tau \right)}{Pe_b}-\frac{k^2 k_a R_a \beta_a (-1+\Gamma_{a,B}) (H_b+\delta ) \tau  \left(\frac{Bo H_b^2}{2}+(H_b+\delta ) \tau \right)}{H_b}\\
		& \nonumber +\mathrm{i} Bo H_b k^3 (H_b+\delta )^2 \tau  \left(\frac{Bo H_b^2}{2}+(H_b+\delta ) \tau \right)+\mathrm{i} k^3 (H_b+\delta )^2 \tau ^2 \left(\frac{Bo H_b^2}{2}+(H_b+\delta ) \tau \right)\\
		& \nonumber -\frac{\mathrm{i} k^5 \left(-3 A+C H_b^4 k^2\right) (2 H_b+3 \delta ) \left(Bo H_b^2+2 (H_b+\delta ) \tau \right)}{12 H_b^2 Pe_b}\\
		& \nonumber +\frac{\mathrm{i} k^3 \left(-3 A+C H_b^4 k^2\right) k_a R_a \beta_a (-1+\Gamma_{a,B}) (2 H_b+3 \delta ) \left(Bo H_b^2+2 (H_b+\delta ) \tau \right)}{12 H_b^3}\\
		& \nonumber +\frac{1}{12} \mathrm{i} Bo^2 H_b^2 k^3 (H_b+\delta ) (2 H_b+3 \delta ) \left(Bo H_b^2+2 (H_b+\delta ) \tau \right)\\
		& \nonumber +\frac{Bo k^4 \left(-3 A+C H_b^4 k^2\right) (2 H_b+3 \delta )^2 \left(Bo H_b^2+2 (H_b+\delta ) \tau \right)}{72 H_b}\\
		& \nonumber +\frac{1}{12} \mathrm{i} Bo H_b k^3 (H_b+\delta ) (2 H_b+3 \delta ) \tau  \left(Bo H_b^2+2 (H_b+\delta ) \tau \right)\\
		& \nonumber +\frac{k^4 \left(-3 A+C H_b^4 k^2\right) (H_b+\delta ) (2 H_b+3 \delta ) \tau  \left(Bo H_b^2+2 (H_b+\delta ) \tau \right)}{12 H_b^2}\Bigg]
	\end{align}
	
	\section*{References}
	\bibliographystyle{unsrtnat}
	\bibliography{Ref}


\end{document}